\documentclass[fleqn,usenatbib]{mnras}

\usepackage{newtxtext,newtxmath}

\usepackage[T1]{fontenc}

\DeclareRobustCommand{\VAN}[3]{#2}
\let\VANthebibliography\thebibliography
\def\thebibliography{\DeclareRobustCommand{\VAN}[3]{##3}\VANthebibliography}

\usepackage{graphicx}	
\usepackage{amsmath}	
\usepackage{hyperref}
\usepackage{diagbox}
\usepackage{xcolor}
\usepackage{multirow, bigdelim}
\usepackage{anyfontsize}

\title[Infall region with UNIONS]{Cosmology from UNIONS weak lensing profiles of galaxy clusters}

\author[C. T. Mpetha et al.]{C. T. Mpetha$^{1,2,3}$\thanks{E-mail: c.mpetha@ed.ac.uk (CTM)},
J. E. Taylor$^{2,3}$,
Y. Amoura$^{2,3}$,
R. Haggar$^{2,3}$,
T. de Boer${^4}$,
S. Guerrini$^{5}$,
\newauthor
A. Guinot$^{6}$,
F. Hervas Peters$^{7}$,
H. Hildebrandt$^{8}$,
M. J. Hudson$^{2,3,9}$,
M. Kilbinger$^{7}$,
T. Liaudat$^{10}$,
\newauthor
A. McConnachie$^{11}$,
L. Van Waerbeke$^{12}$,
A. Wittje${^8}$
\\
$^{1}$Institute for Astronomy, School of Physics and Astronomy, University of Edinburgh,
Royal Observatory, Blackford Hill, Edinburgh, EH9 3HJ, United Kingdom\\
$^{2}$Waterloo Centre for Astrophysics, University of Waterloo, Waterloo, Ontario N2L 3G1, Canada \\
$^{3}$Department of Physics and Astronomy, University of Waterloo, 200 University Avenue West, Waterloo, Ontario N2L 3G1, Canada\\
$^{4}$Institute for Astronomy, University of Hawaii, 2680 Woodlawn Drive, Honolulu HI 96822\\
$^{5}$Universit\'e Paris Cit\'e, Université Paris-Saclay, CEA, CNRS, AIM, F-91191, Gif-sur-Yvette, France\\
$^{6}$Department of Physics, McWilliams Center for Cosmology, Carnegie Mellon University, Pittsburgh, PA 15213, USA\\
$^{7}$Universit\'e Paris-Saclay, Universit\'e Paris Cit\'e, CEA, CNRS, AIM, 91191, Gif-sur-Yvette, France\\
$^{8}$Ruhr University Bochum, Faculty of Physics and Astronomy, Astronomical Institute (AIRUB), German Centre for Cosmological\\
Lensing, 44780 Bochum, Germany\\
$^{9}$Perimeter Institute for Theoretical Physics, 31 Caroline St. N., Waterloo, ON, N2L 2Y5, Canada
$^{10}$IRFU, CEA, Universit\'e Paris-Saclay, F-91191, Gif-sur-Yvette, France\\
$^{11}$NRC Herzberg Astronomy and Astrophysics Research Centre, 5071 West Saanich Road, Victoria, B.C., Canada, V9E 2E7\\
$^{12}$Department of Physics and Astronomy, University of British Columbia, Vancouver, BC V6T 1Z1, Canada
}

\date{Accepted 2025 September 8. Received 2025 August 28; in original form 2025 January 15}

\pubyear{2025}

\begin{document}



\newcommand{\oas}{$\Omega_{\rm m}$ and $\sigma_8$}


\label{firstpage} 
\pagerange{\pageref{firstpage}--\pageref{lastpage}}

\maketitle
\begin{abstract}

Cosmological information is encoded in the structure of galaxy clusters. In Universes with less matter and larger initial density perturbations, clusters form earlier and have more time to accrete material, leading to a more extended \textit{infall region}. Thus, measuring the mean mass distribution in the infall region provides a novel cosmological test. The infall region is largely insensitive to baryonic physics, and provides a cleaner structural test than other measures of cluster assembly time such as concentration. We consider cluster samples from three publicly available galaxy cluster catalogues: the Spectroscopic Identification of eROSITA Sources (SPIDERS) catalogue, the X-ray and Sunyaev-Zeldovich effect selected clusters in the meta-catalogue M2C, and clusters identified in the Dark Energy Spectroscopic Instrument (DESI) Legacy Imaging Survey. Using a preliminary shape catalogue from the Ultraviolet Near Infrared Optical Northern Survey (UNIONS), we derive excess surface mass density profiles for each sample. We then compare the mean profile for the DESI Legacy sample, which is the most complete, to predictions from a suite of simulations covering a range of \oas, obtaining constraints of $\Omega_{\rm m}=0.34\pm 0.06$ and $\sigma_8=0.77 \pm 0.04$. We also measure mean (comoving) splashback radii for SPIDERS, M2C and DESI Legacy Imaging Survey clusters of $1.39^{+0.21}_{-0.18}\,$cMpc$/h$, $1.77^{+0.20}_{-0.18}\,$cMpc$/h$ and $1.42^{+0.11}_{-0.12}\,$cMpc$/h$ respectively. Performing this analysis with the final UNIONS shape catalogue and the full sample of spectroscopically observed clusters in DESI, we can expect to improve on the best current constraints from cluster abundance studies by a factor of $2$ or more.

\end{abstract}




\section{Introduction}
\label{sec:Intro}

Galaxy clusters, the most massive gravitationally bound structures in the Universe, are a versatile probe of fundamental physics. Counting their abundance on the sky is a powerful approach to constraining cosmological parameters \citep{Mantz2015,SPT2024,eROSITA_abundance}. The inner mass profile of clusters has been extensively studied as an alternate test of cosmology \citep[e.g.][]{c_cosmo_1,sparsity_orig}, but uncertain modelling of  baryonic physics is a major limitation. This is particularly significant in the centre of a cluster where powerful but poorly understood AGN feedback mechanisms have a strong impact on the matter distribution \citep[e.g.][]{baryons_central}. This motivates exploring similar structural tests using the outer regions of clusters. Matter in and around a dark matter halo can be separated into two broad categories: \textit{orbiting} or \textit{infalling} \citep{infall_structure}. Orbiting material is gravitationally bound to the halo, and largely lies inside the virial radius. The \textit{infall region} is the zone where the orbiting component is subdominant---most of the material is either falling in for the first time, or is \textit{backsplashed} \citep[e.g.][]{backsplash}, that is approaching or passing through the first apocentre of its orbit after infall. While they currently reside beyond the cluster radius, backsplashed galaxies have previously passed through the cluster centre, and thus have experienced strong environmental effects such as ram-pressure stripping and tidal heating \citep{Borrow2023}. The infall region is typically 1--5\,Mpc from the cluster centre and its upper bound is the turnaround radius, where the expansion of the Universe overcomes the gravitational attraction of the halo. This region is outside the influence of significant baryonic feedback effects \citep{Haggar_2021,Splashback_sims,FLAMINGO,Roan}, so the matter distribution should reflect the assembly history of the cluster. Furthermore, while the difficulty of locating the true cluster centre in observations hampers efforts to use the inner profile, mis-centering has been found to produce differences below the level of the measurement uncertainty in the infall region \citep{sp_obs_1,sp_obs_4,sp_obs_7}. Thus, the infall region provides an interesting environment in which to test for cosmological effects.

There has also been recent theoretical interest in the infall region, as its features may provide a more physical definition of the halo boundary. Classical modelling of the matter distribution around clusters separates
halos (the `1-halo' component of the galaxy clustering signal) from their environment (the `2-halo' component) around the virial radius, usually defined by a mean density contrast relative to the background or critical density. Dark matter halos accrete continuously from their environment, however, and some material beyond the virial radius can have been accreted previously; a more informative halo boundary would extend into the infall region, reflecting this ongoing accretion. Several definitions for this extended halo boundary have been proposed \citep{Diemer14,sp_obs_11,Garcia,FH}, the most notable being the splashback radius $r_{\rm sp}$. This is the radius where infalling material accumulates after it has reached the apocenter of its first orbit, and is sensitive to the net accretion rate over one previous dynamical time \citep{sp_dependence}. 

\cite{Roan} investigated the cosmological dependence of $r_{\rm sp}$, and found its variation with \oas, the matter density fraction and initial amplitude of density perturbations, is nearly orthogonal to $S_8=\sigma_8\sqrt{\Omega_{\rm m}/0.3}$, the parameter best constrained by low-redshift tests of the matter distribution such as cosmic shear or cluster abundance. This cosmological dependence of $r_{\rm sp}$ comes from its relation to cluster assembly times. In cosmologies with low $\Omega_{\rm m}$ and high $\sigma_8$, clusters form earlier \citep{Giocoli2012,Yuba} and have more time to virialise/stabilise by consuming surrounding matter. They provide a deeper potential well for material accreted over the last dynamical time, and thus produce a more extended infall region, and less recent accretion as the halos are already stabilised. This corresponds to a larger splashback radius at late times \citep{Diemer2_cosmo}. The converse is true in cosmologies with high $\Omega_{\rm m}$ and low $\sigma_8$. Given the parameter dependence of cluster formation time is almost orthogonal to that of abundance \citep{Yuba}, observing the infall region could be a highly complementary probe of cosmology to existing low-redshift tests. \cite{Mpetha} investigated the prospects for using weak lensing profiles of galaxy clusters to constrain \oas\ for two weak lensing surveys, \textit{Euclid}-Wide \citep{Euclid_Wide} and the Ultraviolet Near Infrared Optical Northern Survey\footnote{\url{https://www.skysurvey.cc/}} \citep[hereafter UNIONS]{UNIONS_survey}. In particular, they developed a new test based on the amplitude of the shear in the infall region, showing that for the full UNIONS lensing survey, it should produce constraints competitive with those from cluster abundance.

In this work we will conduct the first attempt at such an analysis, comparing the weak lensing profiles of galaxy clusters to a suite of dark matter simulations using different combinations of \oas. We make several improvements to the method presented in \cite{Mpetha}:
including mis-centering in the fitting of weak lensing profiles, performing a Monte Carlo sampling over the fit parameter space, and adding realistic scatter to the masses of simulated halos to reproduce observational selection effects. We explore the challenges of this method, and forecast its potential for present and forthcoming surveys, such as UNIONS combined with a spectroscopic galaxy cluster catalogue of the Dark Energy Spectroscopic Instrument (DESI) survey\footnote{\url{https://www.desi.lbl.gov/}}.

The outline of the paper is as follows. In Section\;\ref{sec:data}, we describe the data products used in this work, including three publicly available cluster catalogues and a weak lensing shape catalogue, that will be combined to construct cluster lensing profiles. Section\;\ref{sec:sims} describes the dark-matter-only cosmological simulation suite we use to generate simulated cluster lensing profiles that will be compared to the observations. Section\;\ref{sec:methods} outlines the main steps in the method, including how excess surface mass density profiles, $\Delta\Sigma(R)$, are determined from the cluster and shape catalogues, how these profiles are compared to the simulation profiles, and finally how profiles are fitted to extract specific features of the infall region, such as the splashback radius $r_{\rm sp}$. Section\;\ref{sec:profile_comparison} presents cosmological constraints derived from the amplitude of the observed $\Delta\Sigma(R)$ in the most complete cluster sample used in this work. In Section\;\ref{sec:fits} we then fit the observed $\Delta\Sigma(R)$ profiles of each sample using a theoretical model of the infall region, determining mean splashback radii for each cluster sample. In Section\;\ref{sec:radii_constraints} we discuss how these fits provide an additional route to constraining cosmological parameters. We discuss our results in Section\;\ref{sec:discussion}, and conclude in Section\;\ref{sec:conclusions}.

The Hubble constant at present is $H_0=100\,h\,$km\,s$^{-1}\,$Mpc$^{-1}$. Throughout this work quantities are scaled by $h^{-1}$, where $h$ is the dimensionless Hubble parameter and its value is fixed at $h=0.7$ when converting from unscaled units. We use various values of $\Omega_{\rm m}$ (as indicated) when computing profiles, such that we can accurately compare observed profiles to profiles in simulations with a chosen $\Omega_{\rm m}$. A prefix of c, for example in cMpc$/h$ denotes a comoving unit, and log is log$_{10}$, while ln indicates log$_e$.

\section{Data}
\label{sec:data}

\begin{figure*}
    \centering
    \includegraphics[width=\linewidth]{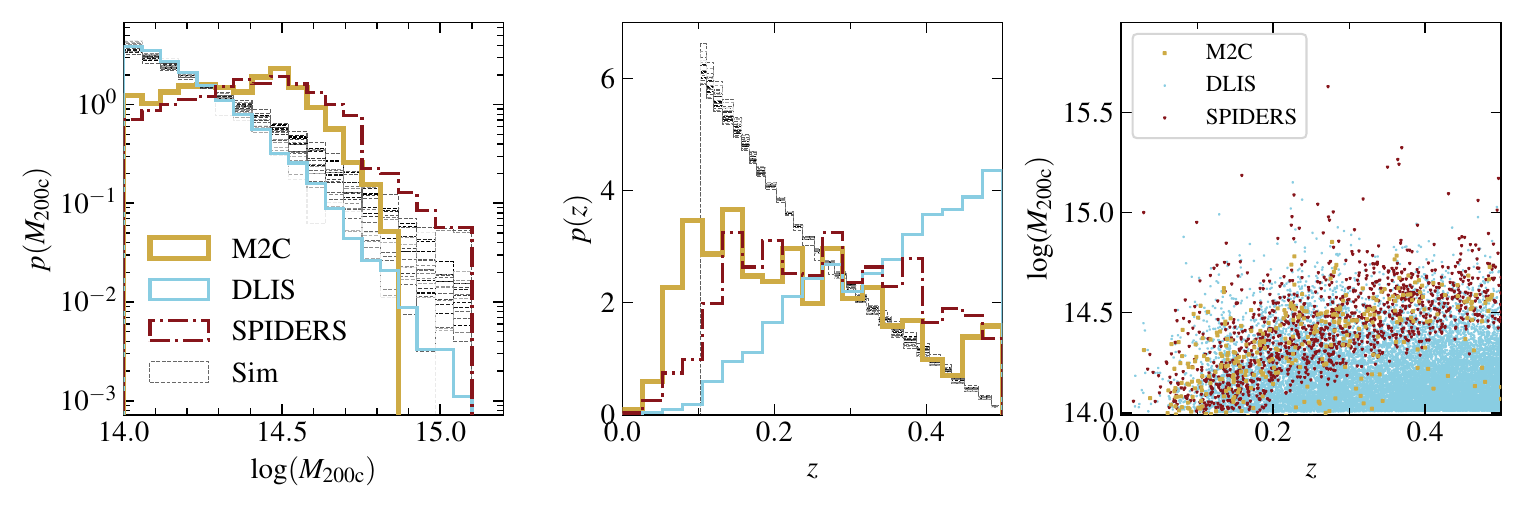}
    \caption{\textit{Left:} Mass distributions of the three cluster catalogues used in this work. M2C and SPIDERS clusters have masses calibrated from an X-ray luminosity -- mass relationship, and in DLIS a total stellar mass-cluster mass relationship is used. For plotting purposes only, $M_{500{\rm c}}$ values in DLIS have been converted to $M_{200{\rm c}}$ using a fiducial cosmology ($\Omega_{\rm m}=0.3$, $\sigma_8=0.8$). Also shown are all the mass distributions of dark matter halos with $0.1<z<0.5$ in $19$ cosmological simulations with different values of \oas, where darker line colours are simulations with larger values of $S_8$. \textit{Middle:} Redshift distributions of clusters. The redshift of halos in the simulations are taken from 29 snapshots between $z=0.1$ and $z=0.5$. The difference in the redshift distributions explains the large discrepancy in the mass function in the left panel. \textit{Right:} The mass-redshift distribution, illustrating the selection biases in the SPIDERS and M2C catalogues.}
    \label{fig:Mz}
\end{figure*}

For the galaxy cluster lensing profiles, we use the weak lensing shape catalogue from the UNIONS survey \citep{shapefit,UNIONS_BH24,Guerrini24}, and publicly available cluster catalogues from the DESI Legacy Imaging Survey \citep[hereafter DLIS]{DESI_Legacy}, the Spectroscopic Identification of eROSITA Sources (SPIDERS) \citep{SPIDERS}, and the M2C cluster catalogue\footnote{\url{https://www.galaxyclusterdb.eu/m2c/}}. These are described in the following sections. The final cluster samples adopted, and the matched samples selected from the simulations, are summarised in Table \ref{tab:lenses}. Normalised mass and redshift distributions, and a mass-redshift scatter plot, are shown in Fig.\;\ref{fig:Mz}. For reference, a plot of the UNIONS footprint and the positions of all clusters in this footprint used in the analysis can be seen in Appendix\;\ref{app:sky}.

\begin{table*}
    \centering
    \begin{tabular}{c|ccccccc}
         \multirow{2}{*}{Catalogue} & \multirow{2}{*}{$N$}& \multirow{2}{*}{$\langle z \rangle$}&log($\langle M_{200{\rm c}} \rangle)$ & $\langle r_{200{\rm c}} \rangle$ & log($\langle M_{200{\rm m}} \rangle)$ &$\langle r_{200{\rm m}} \rangle$ & \multirow{2}{*}{$\nu_{200{\rm m}}$} \\
         & &  &  $M_{\odot}/h$& cMpc$/h$&  $M_{\odot}/h$& cMpc$/h$ & \\
         \hline
    M2C& $384$& $0.236$& $14.52$&  $1.28$& $14.63$&$1.83$&$2.93$\\
    DLIS & $15,\!782$&$0.348$ & $14.21$& $1.06$& $14.31$&$1.43$&$2.60$\\
    DLIS high & $7,\!199$&$0.346$ & $14.33$& $1.16$& $14.42$&$1.56$&$2.76$\\
     SPIDERS &$1,\!225$ & $0.275$& $14.48$& $1.27$& $14.59$&$1.78$&$2.93$\\
     \hline
     Sim& $7,\!782-114,\!341$& $0.276$&$14.20-14.32$& $1.02-1.13$& $14.31-14.43$&$1.43-1.57$&$2.51-2.68$\\
 \end{tabular}
    \caption{For the observed samples, $N$ is number in the UNIONS footprint. $\Omega_{\rm m}=0.3$, $\sigma_8=0.8$ is assumed to convert to $M_{200{\rm m}}$ and the peak height $\nu_{200{\rm m}}$. Radii are given in comoving units. The DLIS high sample uses a larger mass limit. For the simulation halo samples, we give the minimum and maximum value from the full suite.}
    \label{tab:lenses}
\end{table*}

\subsection{UNIONS shape catalogue}
\label{sec:UNIONS}

The Ultraviolet Near Infrared Optical Northern Survey~\citep[UNIONS;][]{UNIONS_survey} is a collaboration of wide field imaging surveys of the northern hemisphere. UNIONS consists of the Canada-France Imaging Survey (CFIS), conducted at the 3.6-meter CFHT on Maunakea, members of the Pan-STARRS team, and the Wide Imaging with Subaru HyperSuprime-Cam of the \textit{Euclid} Sky (WISHES) team. CFHT/CFIS provides deep u and r bands; Pan-STARRS provides deep i and moderate-deep z band imaging, and Subaru provides deep z-band imaging through WISHES and g-band imaging through the Waterloo-Hawaii IfA g-band Survey (WHIGS). These independent efforts are directed, in part, to securing optical imaging to complement the \textit{Euclid} space mission \citep{Euclid2024}, although UNIONS is a separate collaboration aimed at maximizing the science return of these large and deep surveys of the northern skies.

The shape catalogue used in this work was created using the ShapePipe software \citep{shapepipe}. An earlier version of the UNIONS ShapePipe catalogue is described in \cite{shapefit}, and the first use of the version adopted in this work, ShapePipe v1.3, is in \cite{UNIONS_BH24}. There are several improvements in ShapePipe v1.3, including replacing PSFEx \citep{PSFEx} with the Multi-CCD PSF model \citep{MCCD} and reducing contamination by artefacts. ShapePipe v1.3 contains nearly $98\,$million sources over $\sim\!3,\!200\,$deg$^2$, giving a raw number density of $8.5\,$arcmin$^{-2}$. The shapes of source galaxies are measured with the ShapePipe package on $r$-band images taken by MegaCam on CFHT.

\cite{Guerrini24} show first systematics diagnostics for ShapePipe v1.3 and introduce methodologies that will be used in future works to validate the galaxy catalogues used for cosmic shear analysis. In ShapePipe v1.3, there is no estimate for the multiplicative shear measurement bias; however this will be subdominant to other sources of uncertainty and can be safely neglected for our analysis. The sources do not have photometric redshifts currently, but an effective redshift distribution $n(z)$ has been constructed using Self Organising Maps (SOMs) \citep{SOM}. UNIONS galaxies in the CFHTLenS \citep{CFHTLens} field W3 are chosen as a representative subset in order to utilise the colour information of the ugriz photometry from CFHTLenS \citep{CFHTLens_phot}. The SOM $n(z)$ for these galaxies is calibrated with the colours of $\sim\!66,\!000$ galaxies from the deep spectroscopic surveys DEEP2 \citep{DEEP2}, VVDS \citep{VLT}, and VIPERS \citep{VIPERS} as reference samples. This $n(z)$ is assumed to trace the true distribution of the full sample. More detail can be found in Appendix\;A of \cite{UNIONS_BH24}. Both simulation-based tests and bootstrap resampling estimation of the uncertainty in the SOM method found the bias in $n(z)$ and its uncertainty to be well within expectations for Stage III surveys. Thus, we expect a bias on $n(z)$ to have a negligible impact on our analysis.

\subsection{DESI Legacy Imaging Survey cluster catalogue}
\label{sec:DESI_Legacy}

\cite{DESI_Legacy} present a catalogue of $1.58\,$million galaxy clusters from the DESI Legacy Imaging Survey (DLIS)\footnote{\url{http://zmtt.bao.ac.cn/galaxy_clusters/catalogs.html}}. Of these, $178,\!788$ lie in the UNIONS footprint. To identify clusters, the authors first find a BCG candidate catalogue from the initial DESI Legacy galaxy catalogue based on stellar mass and colour. Some BCGs have a spectroscopic redshift, in which case the cluster redshift is well measured. In some cases, one or more identified cluster members within $|\Delta v| <2500\,$km$\,$s$^{-1}$ of a BCG have spectroscopic redshifts, and these are used to estimate the redshift of the cluster. In many cases, only photometric redshifts are available. We only use systems with a spectroscopically estimated redshift of either the BCG or a possible cluster member, which reduces the number to $89,\!913$ clusters. This sample is still likely to have interloper clusters caused by projection effects. The spectroscopic redshift selection has a small impact on the cluster mass function, and for the redshift distribution it retains the general trend while removing spurious features. We conclude that only using clusters with at least one spec$-z$ does not induce any further selection effects on the sample.

In \cite{DESI_Legacy} the cluster mass is determined from a total stellar mass-cluster mass relationship. First, they find the stellar mass in a radius $(H(z)/H_0)^{-2/3}\,{\rm Mpc}=(0.7+0.3(1+z)^3)^{-2/3}\,$Mpc around the BCG. This is related to $r_{500{\rm c}}$ through a scaling relation measured from a large set of clusters \citep{Wen2015}, which is rescaled using the results of \citet{Vikhlinin2009} as their mass estimates show $\lesssim10\%$ difference compared to weak lensing masses. From $r_{500{\rm c}}$, they measure the total stellar mass within $r_{500{\rm c}}$, and then obtain $M_{500{\rm c}}$ from a scaling relation between the stellar to cluster halo mass within that radius, which is described in \cite{Wen2021}. We further limit the sample to clusters with $M_{500{\rm c}} > 10^{13.85}M_{\odot}/h$ (corresponding closely to $M_{200{\rm c}} > 10^{14}M_{\odot}/h$) and $z\leq0.5$. We also define a sample with a larger mass limit of $M_{500{\rm c}} > 10^{14}M_{\odot}/h$ (DLIS high); this higher mass limit removes many spurious systems. The redshift limit ensures all of the cluster BCGs have a $z-$band magnitude $m_z < 21$, which from Fig.\;1 of \cite{DESI_Legacy} significantly reduces the bias and scatter of photometric redshift estimates. Also, given the $n(z)$ of the UNIONS shape catalogue peaks at $z\!\sim\!0.6$, the inclusion of higher-redshift clusters will mostly contribute noise. The final number after imposing these constraints is $15,\!782$. The quoted mass uncertainty of these clusters is $0.2\,$dex. 

The authors state their halo sample demonstrates good completeness. We do not expect their halo mass function to be similar to those in the simulations, however, due to the different redshift distributions (see Fig.\;\ref{fig:Mz}).

\subsection{SPIDERS cluster catalogue}
\label{sec:SPIDERS}

SPIDERS \citep{SPIDERS} is the Sloan Digital Sky Survey IV \citep[SDSS-IV,][]{SDSSIV} spectroscopic follow up of a subset of CODEX galaxy clusters \citep{CODEX}, originally identified in the ROSAT All-Sky X-ray Survey. RedMaPPer \citep{redmapper_orig} is used to identify potential cluster members for spectroscopic follow-up in SDSS imaging data. Follow-up is performed with the Baryon Oscillation Spectroscopic Survey spectrograph \citep{BOSS}, producing spectroscopic membership, accurate cluster redshifts, and cluster velocity dispersions. The cluster mass is iteratively calibrated from X-ray luminosity -- mass, mass -- temperature and luminosity -- temperature relationships \citep{CODEX}, producing results consistent with the CODEX weak lensing mass calibration of \cite{Kettula2015}. The mean mass uncertainty of SPIDERS clusters is $0.2\,$dex.

In total there are $2,\!740$ clusters. Of these, $1,\!576$ lie in the UNIONS footprint. We then restrict the sample to clusters with $M > 10^{14}\,M_{\odot}/h$, and $z\leq0.5$. We also impose \texttt{NCOMPONENT}$\,=1$ to remove merging systems from our analysis. The final number is $1,\!213$. Since the cluster sample is cross-matched between both an X-ray survey and a spectroscopic galaxy catalogue, systems are much less likely to be caused by projection effects than in the DLIS catalogue.

The SPIDERS catalogue includes centroids determined from the ROSAT X-ray image, as well as centroids determined by redMaPPer. Given the large centering uncertainties in low signal-to-noise ratio (SNR) X-ray data, we adopt the more reliable optical centroids. Comparing the mean lensing profiles derived for either choice of centroid, we find the signal is stronger towards the centre when the optical centroids are used, confirming that these are more accurate. 

There are significant possible selection biases associated with using this catalogue for cosmology. First, the sample is not complete, with the mass distribution deviating considerably from the halo mass function. This means we must be careful in how we compare the sample to simulations. Second, X-ray detected clusters are more likely to be dynamically relaxed, as this will produce a more concentrated core and a higher peak of X-ray emission \citep{Xray_bias}. Dynamically relaxed clusters will have a more extended infall region, biasing cosmological inferences. \cite{Diemer2_cosmo} use simulations to demonstrate that dynamically active halos with larger accretion rates have up to $\sim\!50\%$ smaller splashback radii (see their Fig.\;6). Given the smaller sample size in SPIDERS and these selection effects, we do not attempt to derive cosmological constraints with this sample.

\subsection{M2C cluster catalogue}
\label{sec:M2C}

The M2C galaxy cluster database is a combination of three catalogues: MCXC, a meta catalogue of X-ray detected clusters in Einstein and ROSAT surveys \citep{MCXC,MCXC-II}; MCSZ, a meta catalogue of Sunyaev–Zeldovich (SZ) effect detected clusters in \textit{Planck} \citep{M2C_Planck_1,M2C_Planck_2,M2C_Planck_3}, SPT \citep{M2C_SPT_1,M2C_SPT_2} and ACT \citep{M2C_ACT_1,M2C_ACT_2}; and ComPRASS, joint X-ray/SZ detected clusters in RASS and \textit{Planck} surveys \citep{comprass}. Masses are determined using X-ray luminosity -- mass, or (X-ray calibrated) SZ signal -- mass relationships. As these masses are not calibrated with weak lensing measurements, they could suffer from a hydrostatic mass bias \citep{Hurier2017,Braspenning2024}. The catalogue contains $5,\!452$ unique entries in total, including new objects in MCXC-II. Of these, $576$ have an associated mass and lie in the UNIONS footprint. Imposing mass and redshift cuts of $M > 10^{14}\,M_{\odot}/h$, and $z\leq0.5$, reduces the final number to $385$. 

The updated MCXC-II catalogue \citep{MCXC-II} has an average fractional uncertainty of $\langle\Delta M_{500{\rm c}}/M_{500{\rm c}}\rangle = 0.12$. In MCSZ, the mean fractional mass uncertainty is $\sim\!0.14\,$dex, while in ComPRASS it is $0.2\,$dex. We note that there is considerable overlap between the M2C and SPIDERS catalogues; using TopCat \citep{Topcat} we identified $686$ likely matches, assuming a maximum redshift separation of $0.025$. Thus in what follows, we do not treat these as completely independent datasets. We expect the M2C catalogue to have selection biases similar to those in the SPIDERS sample, though even more pronounced, given the smaller number of sources.

\section{Simulations}
\label{sec:sims}

We compare the mean projected mass profiles of the observed clusters to those of dark matter halo samples drawn from a suite of $19$ dark-matter-only simulations with a grid of values of \oas, described in \cite{Yuba_thesis}. These cover the range $0.2\leq\Omega_{\rm m}\leq0.4$ and $0.7\leq\sigma_8\leq1$; the exact combinations are shown in Table\;1 of \cite{Roan}. Other parameters in the simulations are fixed, including a Hubble parameter $H_0 = 100\,h = 70\,$km$\,$s$^{-1}\,$Mpc$^{-1}$, a baryon density $\Omega_{\rm b}=0.0482$, and a spectral tilt $n_s=0.965$. The simulations were created using \textsc{Gadget 4} \citep{Gadget4} and contain $1024^3$ particles in a $500\,$Mpc$/h$ box, giving a particle mass of $\Omega_{\rm m}\times (3.23\times10^{10}\,M_{\odot}/h)$. The set of initial conditions of each simulation is the same, such that variations in the mean halo profiles reflect cosmology rather than cosmic variance. The suite will be described in more detail in a forthcoming work (Amoura et al.~2025, in preparation). 

Given the fairly sparse sampling in \oas\ space, we estimate that constraints on these parameters derived from our analysis will have a precision of $\pm0.025$ at best (half the step-size in cosmological parameters). A natural extension to the work presented here would be to repeat our analysis using a finer grid of cosmologies, also exploring the effect of varying other cosmological parameters.

The \textsc{Amiga Halo Finder} (AHF) \citep{AHF} was used to identify halos in the 119 snapshots output from each simulation. AHF employs Adaptive Mesh Refinement (AMR) to split the simulation volume into progressively smaller grids in regions of high particle density, and then searches for halos, starting with the smallest grids. Subhalos within larger halos are identified using the AMR grid-tree structure (see Fig.\;1 of \cite{AHF}); we remove these from the final halo sample. Halo centres and virial radii $r_{200{\rm c}}$, are found from (particle) iso-density contours. 

We use a set of cluster-mass halos from the 29 snapshots between $z=0.1$ and $z=0.5$, providing a mean redshift similar to the observed samples from the M2C and SPIDERS catalogues. Though the redshift extent is the same, the redshift distributions of the simulated and catalogue samples differ, most notably for DLIS, having a greater number at low redshift. To address this, when creating simulation profiles to compare with the DLIS profiles, we will re-weight the contribution of a simulated halo based on the ratio of the number of observed clusters and simulated halos at that halo's redshift. 

From the simulated samples, we determine the mean projected excess mass profile $\Delta\Sigma(R)$, as this corresponds most closely to the quantity that can be determined from observations \citep{Diemer25}.

\section{Methods}
\label{sec:methods}

\subsection{Lensing profiles from observations}

\begin{figure*}
    \centering
    \includegraphics[width=\linewidth]{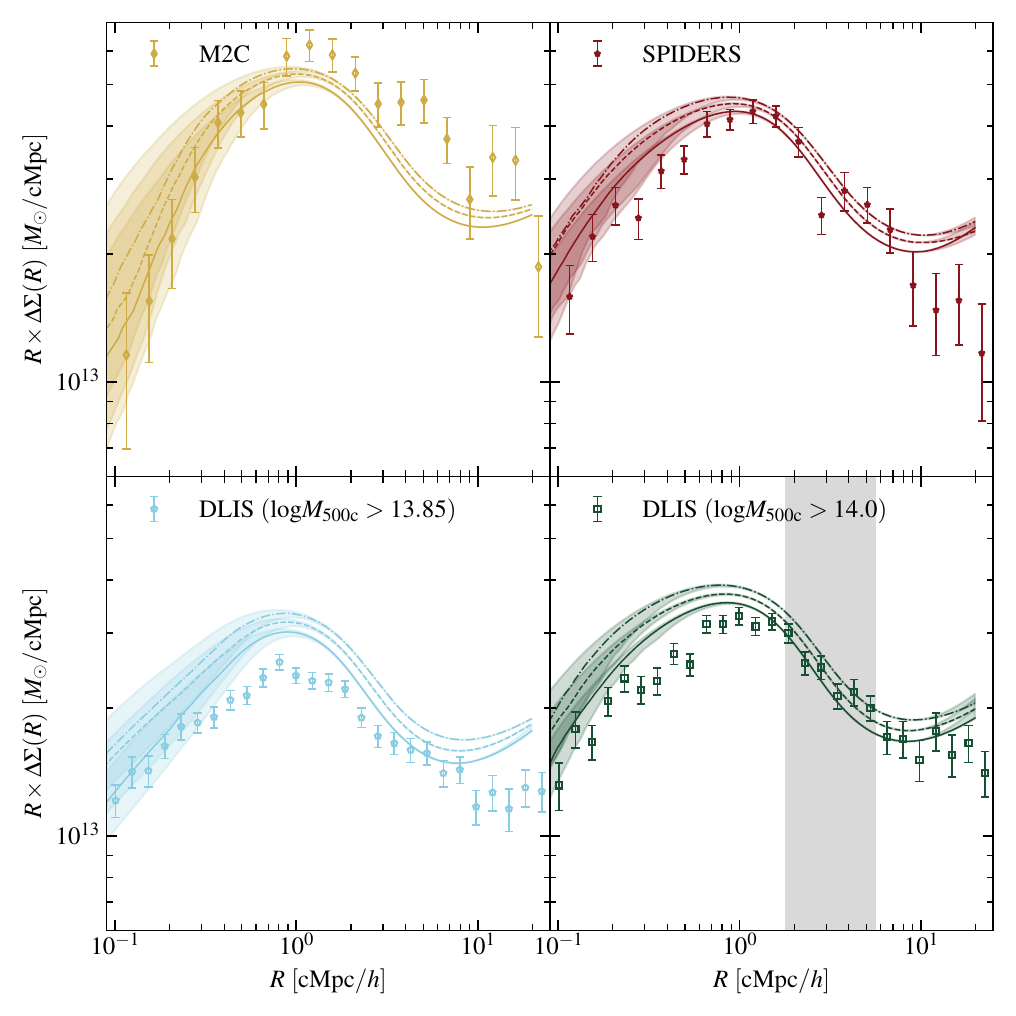}
    \caption{Data points with error bars show mean excess mass density profiles $\Delta\Sigma(R)$ (in comoving units) for three publicly available lens catalogues, calculated with the code \textsc{dsigma} from a UNIONS shape catalogue. The DESI Legacy Imaging Survey (DLIS) catalogue is used with two choices for the minimum mass in the subsample (bottom panels). The value $\Omega_{\rm m}=0.3$ is used to convert angles and redshifts to distances. Also overlaid are the corresponding mean profiles from simulations with $\Omega_{\rm m}=0.3$ and $\sigma_8=0.7, 0.8, 0.9$ (solid, dashed, dash-dot respectively). The uncertainty in a simulation profile is given by a shaded band. For DLIS, the match is performed using all simulated halos above the quoted mass limit and weighting the contribution of each halo to the mean profile based on the ratio of observed clusters and simulated halos at that halo's redshift. For M2C and SPIDERS, when taking the mean we weight the contribution of each simulated halo by the relative number of halos at that mass in the observed sample. We have corrected the simulated profiles for mis-centering, based on fits to the observed data. The grey band in the bottom right panel indicates the radial range used to compare the observed profile to the simulated profiles in our cosmological analysis.}
    \label{fig:obs_profiles}
\end{figure*}

The observable in cluster lensing is the excess surface mass density, $\Delta\Sigma$. It can be expressed as the difference between the average surface mass density within a projected radius $R$, and the mass density at that radius:
\begin{equation}
        \Delta \Sigma(r) = \overline{\Sigma}(R)-\Sigma(R) \, ,
        \label{eq:dSigma}
    \end{equation}
where
\begin{equation}
        \overline{\Sigma}(R) = \frac{2}{R^2} \int_0^R R' \Sigma(R') \, dR' \, ,
    \end{equation}
and the surface mass density is a 2D projection of the 3D density profile $\rho(r)$:
 \begin{equation}
        \Sigma(R) = 2\int_R^\infty \frac{\rho(r)r}{\sqrt{r^2 - R^2}}\, dr \, .
        \label{eq:sigma}
    \end{equation} 
The $\Delta\Sigma$ profile of a foreground lens is simply related to the tangential shear $\gamma_{\rm t}$ of background sources through
\begin{equation}
    \Delta\Sigma(R) = \gamma_{\rm t}(R) \Sigma_{\rm crit} \, ,
\end{equation}
where $\Sigma_{\rm crit}$ is the critical density defined below, that depends on the source and lens redshifts. In practice we measure the tangential shape, $e_{\rm t}$, of many background sources around a large number of lenses, and stack the measurements, with appropriate weights, into a composite $\Delta\Sigma$ profile. The composite profile is calculated by summing over lens-source pairs (l--s) in radial bins, 
\begin{equation}
\Delta\Sigma(R) = {\rm boost}(R) \times \frac{\sum_{\rm l-s}w_{\rm sys, l}w_{\rm s} e_{\rm t} \langle \Sigma_{\rm crit, l}^{-1}\rangle}{\sum_{\rm l-s}w_{\rm sys, l}w_{\rm s}\langle \Sigma_{\rm crit, l}^{-1}\rangle^2} \, , \label{eq:esd}
\end{equation}
where $w_{\rm s}$ is the source weighting related to the shape noise, $w_{\rm sys, l}$ is a lens weighting chosen to reduce the impact of lens selection biases, and 
\begin{align}
    \Sigma_{\rm crit, l-s} &= \frac{c^2}{4\pi G}\frac{\chi_{\rm s}}{\chi_{\rm l}(\chi_{\rm s}-\chi_{\rm l})(1+z_{\rm l})} \, ,\\
    \langle \Sigma_{\rm crit, l}^{-1}\rangle &= \int \Sigma_{\rm crit, l-s}^{-1} n(z_{\rm s}) \, dz_{\rm s} \, . \label{eq:sigmacrit}
\end{align}
The $\chi$ indicate comoving distances. For the public lens catalogues adopted, we set $w_{\rm sys,l}=1$ for all sources. Instead of using the inverse 
$\Sigma_{\rm crit}$ value for each lens-source pair, we calculate the average defined above, as the version of the ShapePipe catalogue used in this work does not have photo$-z$'s associated with individual sources. The mean inverse $\Sigma_{\rm crit}$ in Eq.\;(\ref{eq:sigmacrit}) is calculated using the distribution of galaxies in a matched redshift distribution $n(z)$, detailed in Section\;\ref{sec:UNIONS}.

The boost term comes from contamination of the source galaxy sample by cluster members that do not contribute to the lensing signal. This reduces the mean value of the shear, and needs to be corrected for when estimating $\Delta\Sigma$. The boost factor in a radial bin can be estimated by comparing lens-source pairs to random-source pairs,
\begin{equation}
    {\rm boost}(R) = \frac{\sum_{\rm l-s}w_{\rm sys, l}w_{\rm l-s}}{\sum_{\rm r-s}w_{\rm sys, l}w_{\rm l-s}} \, .
    \label{eq:boost}
\end{equation}

We compute mean $\Delta\Sigma$ profiles for the three public galaxy cluster catalogues; M2C, SPIDERS and DLIS, using UNIONS galaxy shapes and the \textsc{dsigma} package \citep{dsigma}. In Appendix\;\ref{sec:boost} we show the size of the boost factor correction for each cluster sample used in this work.

It is important to note that we are assuming $\langle e_{\rm t} \rangle = \gamma_t$, i.e. that the average ellipticity of a large number of galaxies is equal to the tangential shear. The true observable is in fact the reduced shear,
\begin{equation}
    g_{\rm t} = \frac{\gamma_{\rm t}}{1-\kappa} \, .
\end{equation}
The convergence $\kappa$, proportional to the surface mass density, is small at large projected radii, but increases towards the centre of a cluster. The error incurred in assuming $g_{\rm t} = \gamma_{\rm t}$ (i.e.~$\kappa \ll 1$) will be less than $1\%$ in the infall region, and thus will not impact our cosmological constraints. But assuming $g_{\rm t} = \gamma_{\rm t}$ leads to an overestimate of $\Delta\Sigma$ in the inner region, which will affect fits to the inner profile. In Appendix\;\ref{app:kappa} we show that $\kappa$ falls below $20\%$ in the radial range adopted for our fits, indicating the reduced shear approximation will have a small impact on the profile fits.

In addition to the boost correction based on random-source pairs, \cite{Singh17} show that subtracting a $\Delta\Sigma$ profile calculated from random-source pairs (without any boost correction) leads to a cleaner profile and smaller uncertainties. To apply this correction, we generate a set of random locations $20\times$ larger than the lens sample size, drawn uniformly from the spatial distribution traced by lenses within the UNIONS footprint, and assign random redshifts drawn from the same redshift distribution as the lenses. We calculate the resulting $\Delta\Sigma$ profile from Eq.\;(\ref{eq:esd}), and subtract it from the observed profile. We confirm that both the signal from randoms, and the cross component of the signal $\Delta\Sigma_{\times} = \gamma_{\times}\Sigma_{\rm crit}$, are unbiased and consistent with zero.

To calculate the uncertainty, a jackknife resampling method is used. The region is split into $N_{\rm jk}=100$ sub-regions of roughly equal size, then the covariance between radial bins $i$ and $j$ is given by
\begin{equation}
    {\rm Cov}_{ij} = \frac{N_{\rm jk}-1}{N_{\rm jk}} \sum_{k=1}^{N_{\rm jk}} \left(\Delta\Sigma_k(R_i) -\Delta\Sigma(R)\right) \left(\Delta\Sigma_k(R_j) -\Delta\Sigma(R)\right) \, .
\end{equation}

We also tested the impact of a \textit{lens magnification bias}, when lenses are themselves magnified by foreground structure which can contribute to the shear signal. This is more impactful for galaxy-galaxy lensing. Even when assuming extreme responses of lenses to magnification, values, up to $\alpha_{\rm lens}=10$ \citep{lensmag}, the impact on the recovered $\Delta\Sigma$ is $\lesssim 1\%$, such that we can safely neglect this effect.

We calculate $\Delta\Sigma(R)$ using $20$ logarithmically spaced radial bins from $0.1-25\,$cMpc$/h$ for SPIDERS and M2C, and $30$ bins for DLIS. Fig.\;\ref{fig:obs_profiles} shows the resulting profiles for each of the cluster samples. We note that the conversion from angles and redshifts to distances is affected by the choice of $\Omega_{\rm m}$. In Fig.\;\ref{fig:obs_profiles} we have set $\Omega_{\rm m}=0.3$. 
We test the correlation between radial bins from the mean off-diagonal correlation in the correlation matrix (removing the on-diagonal unity terms, and finding the average). This gives $0.13$, $0.16$, $0.09$ and $0.10$ for SPIDERS, M2C, DLIS and DLIS high respectively, indicating little correlation between bins.

\subsection{Comparison to simulations}
\label{sec:simulation_comparison}

The mean density profile from simulated halos, and a corresponding $\Delta\Sigma$ profile is calculated. First the 2D surface mass density as a function of projected cluster-centric distance $\Sigma(R)$, shown in Eq.\;(\ref{eq:sigma}), is found from the density extracted from particle data. Then a corresponding $\Delta\Sigma$ profile is found using Eq.\;(\ref{eq:dSigma}). When computing a final $\Delta\Sigma$ profile from observations, values for $H_0$ and $\Omega_{\rm m}$ are needed to convert angles and redshifts into radii and distances. The value of $H_0=70\,$km$\,$s$^{-1}\,$Mpc$^{-1}$ used in the simulations is kept fixed throughout. As we are comparing to simulations with five different possible values for $\Omega_{\rm m}$, five $\Delta\Sigma$ profiles are calculated for each cluster sample. Then when comparing observations to simulations, profiles with $\Omega_{\rm m}=0.3$, for example, will be compared to the simulations with $\Omega_{\rm m}=0.3$ and $\sigma_8=0.7, 0.75, 0.8, 0.85, 0.9$ to find the best fit.

To obtain accurate cosmological constraints, we need to consider carefully the sample selection for both the observed clusters and the simulated halos. Ideally, the observed cluster samples would be pure (containing only systems over some mass) and complete (containing all systems over that mass); then we would compare these to a mass-limited sample of simulated halos. Uncertainties in cluster masses complicate the situation, however, scattering clusters above the mass cut out of the sample and clusters under the mass cut into the sample --- an Eddington bias. Given the steep slope of the mass function, the dominant effect of errors in mass is to add low-mass systems to the sample.

Another complicating effect is that of mis-centering. Although mis-centering causes a shift smaller than our measurement uncertainties in the infall region, we still include the effect for visual comparison between simulated and observed halos in the inner region, and for accurate profile fitting. There may also be systematic offsets in the masses recovered from the lensing profiles and the catalogue masses determined from observational proxies such as X-ray luminosity. We will consider these effects in future work.

A richness cut is another approach to filtering the observed cluster sample. This would require populating the dark matter only simulations with galaxies, for example using a Halo Occupation Distribution model \cite[e.g.][]{Peacock2000}, such that the same richness cut can then be applied in the simulation sample. This approach trades uncertainty in the mass-observable relation used to derive cluster masses with uncertainty in how galaxies populate dark matter halos. For simplicity, we use the mass-observable relation in this work, but will investigate other approaches in future works.

For each of the cluster samples, we generate a matched profile from the simulations, also including the effects of mass scatter and mis-centering. Projection effects are not included in the simulation profiles --- we will investigate this further in a future work. To make the corresponding simulation profile for the DLIS sample, we use the concentration values for each simulated halo identified by AHF to convert their masses from $M_{200{\rm c}}$ to $M_{500{\rm c}}$, in line with the mass definition used in the DLIS catalogue. When iterating over each halo in the simulation, first an individual mass-scatter is added by drawing from a log-normal distribution with a width equal to the quoted mass uncertainty (in dex units) in the corresponding observed catalogue. A density profile is found by taking the mean profile of the mass-scattered halos, only using halos with a scattered mass above the chosen mass threshold. Then, drawing random values for the fraction of sources with an offset and the amplitude of the offset distribution based on their mean and uncertainty from the corresponding best-fits (see Section\;\ref{sec:fitting}), we create a mis-centered $\Delta\Sigma$ profile. This is repeated $100$ times, and the average of this profile and the $1\sigma$ spread of all the profiles is used as the corresponding simulated profile and its uncertainty. The resulting uncertainties, seen as shaded bands in Fig.\;\ref{fig:obs_profiles}, are larger than those coming from a bootstrap process when creating the density profile from the collection of halos, and so we use these uncertainties when comparing between simulated and observed profiles.

To compare observed and simulated samples, we need to account for completeness in the former. Considering Fig.\;\ref{fig:Mz}, only the DLIS catalogue is reasonably complete, though with a different redshift distribution. Our approach to creating a reliable matched mean simulation profile for DLIS is to re-weight the contribution of each simulation halo to the mean profile based on the ratio between the number of catalogue clusters and simulation halos at that halo's redshift. To create matched simulation profiles for SPIDERS and M2C shown in Fig.\;\ref{fig:obs_profiles}, we weight the contribution of each simulated halo by the ratio of the observed and simulated mass functions. The incompleteness of these catalogues nonetheless suggests they may be prone to selection biases related to dynamical state, so we do not use these samples to derive cosmological constraints.

The lines in Fig.\;\ref{fig:obs_profiles} show the $\Delta\Sigma$ profiles of the matched halo samples in the suite of simulations, corresponding to the observed cluster samples used in this work and outlined in Table\;\ref{tab:lenses}. Each line is from a simulation with $\Omega_{\rm m}=0.3$, and $\sigma_8=0.7,0.8,0.9$ given by the solid, dashed, and dashed dot lines. M2C exhibits excess signal in the infall region, at distances of $2-3\,$cMpc/$h$ (top left panel). This may indicate the catalogue is biased towards dynamically relaxed systems with a larger peak of X-ray emission, which have more extended infall regions \citep{Haggar2020}. For the DLIS sample with log$M_{500{\rm c}}<13.85$ (bottom left panel), the signal is considerably weaker than predicted in the simulations. This is likely due to the inclusion of spurious systems caused by projection effects and large redshift uncertainties of photometrically identified cluster members. We also show a DLIS profile with a higher mass threshold, `DLIS high', along with predictions for simulated samples with the same mass range (bottom right panel). This leads to a much better agreement between the observed and simulated profiles. In both SPIDERS and DLIS high, the observed signal in the inner region appears suppressed compared to the prediction from simulations. This could indicate there is a stronger degree of mis-centering in the observed clusters than we have included in the simulation profiles. The SPIDERS sample shows good agreement with the matched simulation profiles, though with larger errors and scatter compared to DLIS due to the smaller number of systems.

\subsection{Fitting procedure}
\label{sec:fitting}

\begin{table*}
    \centering
    \begin{tabular}{c|lcll}
         Parameter &Fiducial value&  Prior &Description\\
         \hline
         log$\,\rho_s\,/\,[h^2 M_{\odot}/{\rm cMpc}^3]$ &log$(10^3 \, \rho_{\rm m})$& 
     log$([10^1,10^7] \, \rho_{\rm m})$&Density at scale radius & \rdelim\}{3}{*}[Einasto] \\ 
 log$\,r_s\,/\,[$cMpc$/h]$ &log$(0.07\,r_{200{\rm m}})$& log$([0.01,0.45]\,r_{200{\rm m}})$&Scale radius& \\
 log$\,\alpha$ &-1& log$([0.03,0.4])$&Slope of inner Einasto profile& \\ 
 log$\,r_t\,/\,[$cMpc$/h]$ &log$(r_{200{\rm m}})$& log$([0.455,3]\,r_{200{\rm m}})$&Truncation radius for inner Einasto profile & \rdelim\}{2}{*}[Truncation]\\ 
 log$\,\beta$ &0& $[-1,1.3]$&Sharpness of truncation & \\ 
 log$\,\delta_1$ &1& $[0,2]$&Overdensity at the pivot radius & \rdelim\}{3}{*}[Infalling] \\
 log$\,\delta_{\rm max}$ &2& $[0,3+{\rm log}(2)]$&Overdensity in the halo centre& \\
 log$\,s$ &-1& $[-2,2{\rm log}(2)]$&Slope of the infalling term & \\ 
  log$\,\sigma_{\rm off}\,/\,[$cMpc$/h]$&-0.5& $[-2,0]$&Amplitude of offset distribution & \rdelim\}{2}{*}[Mis-centering] \\
 $f_{\rm off}$ &0.3& $[0,1]$&Fraction of lenses with offset& \\\end{tabular}
    \caption{Parameters used to fit a $\Delta\Sigma$ profile. The reference value of $\rho_{\rm m}$ and $r_{200{\rm m}}$ are calculated at $z=0$ as we adopt comoving units, and using the same $\Omega_{\rm m}$ value as used when calculating the $\Delta\Sigma$ profile from the observed shear profile.}
    \label{tab:params}
\end{table*}

We fit the mean observed excess mass profiles using the 3D density profile model of \cite{Diemer_fit} (the basic method in obtaining this profile is outlined in \cite{Diemer2022I}), as this includes an infalling term not present in the Navarro-Frenk-White (NFW) \citep{NFW} or Einasto profiles \citep{Einasto}. This model is designed to be a good fit, with $\lesssim 5\%$ accuracy, to mean dark matter density profiles over a wide range of mass, redshift and cosmology. The model represents an improvement over an earlier version \citep{Diemer14}, which has been used in the literature to estimate the splashback radius of stacked clusters. The new model has been shown to outperform the earlier model and be more physically motivated \citep{Diemer_fit}. It is given by
\begin{align} 
        \rho(r) &= \rho_{\rm orbit}(r) + \rho_{\rm infall}(r) \, ,\\
        \rho_{\rm orbit}(r) &= \rho_s \, \exp{\left(-\frac{2}{\alpha}\left[ \left(\frac{r}{r_s}\right)^{\alpha}-1\right]-\frac{1}{\beta}\left[\left(\frac{r}{r_t}\right)^{\beta}- \left(\frac{r_s}{r_t}\right)^{\beta}\right]\right)}  \nonumber \, ,\\
         \rho_{\rm infall}(r) &= \rho_{\rm m}\,\left(1+ \delta_1/\sqrt{\left(\delta_1/\delta_{\rm max}\right)^2+\left(r/r_{\rm pivot}\right)^{2s}}\right) \, ,
         \label{eq:rho_model}
    \end{align} 
    where
    \begin{equation}
        r_{\rm pivot}(z) = \left(\frac{3 M_{200{\rm m}}}{4\pi \times 200\,\rho_{\rm m}(z)}\right)^{1/3} \, .
    \end{equation}
When fitting simulated density profiles, the parameter $\rho_{\rm m}$ is left free to minimize the residuals. When fitting observed $\Delta\Sigma$ profiles, $\rho_{\rm m}$ is fixed to the matter density using the same assumed value of $\Omega_{\rm m}$ which converts angles and redshifts to distances. The values for $M_{200{\rm m}}$ listed in Table\;\ref{tab:lenses} are used to calculate the pivot radius, but the fits are fairly insensitive to this choice.

The procedure outlined in \cite{Diemer_fit} Appendix\;A is used to find an initial fit to both the observed $\Delta\Sigma$ profiles, and the simulated density profiles. When fitting $\Delta\Sigma$, we start by generating a density profile using the model of Eq.\;(\ref{eq:rho_model}) and a set of model parameters, and then calculate the associated $\Delta\Sigma$ profile. As pointed out in \cite{Diemer25}, values for $\alpha$ and $\delta_{\rm max}$ can be fixed when fitting mean density profiles over the whole radial range, leading to narrower and unbiased posteriors for the remaining parameters. For the observed $\Delta\Sigma$ profiles, we first perform the full least-squares regression procedure, varying all parameters. We then use the Preconditioned Monte Carlo (PMC) algorithm, \textsc{pocoMC} \citep{pocomc1,pocomc2}, to sample from the correlated, non-Gaussian parameter space and obtain uncertainties on the fitting parameters. The initial state of the PMC is the best-fit model from the least-squares regression, and the prior comes from combining the Gaussian uncertainty from the least-squares fit with the uniform prior in Table\;\ref{tab:params}. Finally we repeat this procedure, fixing log$\alpha$ and log$\delta_{\rm max}$ to their maximum a posteriori values from the initial MC chain. We include a mis-centering term into the fitting procedure following e.g.~\cite{Johnston07}. It modifies the profile by constructing it as the sum of a term with no offset, and a term where the centre is mis-identified:
\begin{equation}
    \Sigma(R) = (1-f_{\rm off})\Sigma_0(R) + f_{\rm off}\Sigma_{\rm off}(R) \, ,
\end{equation}
where $f_{\rm off}$ is the unknown fraction of sources with an offset. $\Sigma_0$ is the term in Eq.\;(\ref{eq:sigma}). The offset surface density is given by
\begin{align}
    \Sigma_{\rm off}(R|R_{\rm off}) &= \frac{1}{2\pi} \int_0^{2\pi} \Sigma_0\left(\sqrt{R^2 + R_{\rm off}^2 + 2RR_{\rm off}\cos(\theta)}\right) \,  d\theta \, , \\
    \Sigma_{\rm off}(R) &= \int_0^{\infty}p(R_{\rm off}) \Sigma_{\rm off}(R|R_{\rm off}) \, dR_{\rm off} \, .
\end{align}
The offset probability follows a Rayleigh distribution,
\begin{equation}
    p(R_{\rm off}) = \frac{R_{\rm off}}{\sigma_{\rm off}^2} \exp\left(-\frac{R_{\rm off}^2}{2\sigma_{\rm off}^2}\right) \, .
\end{equation}
The amplitude of the offset distribution, $\sigma_{\rm off}$, has units of cMpc$/h$. Then there are two more free parameters in the fitting, $f_{\rm off}$ and $\sigma_{\rm off}$. Table\;\ref{tab:params} summarizes the fit parameters and their priors, and gives a brief description of each.

We calculate the average, upper and lower bounds for $r_{\rm sp}$ and $\Delta\Sigma$ by recomputing their values for each parameter set in the final MC chain, and finding the median and $68\%$ confidence interval.

\section{Results}
\label{sec:results}

Our analysis proceeds through three main steps:
\begin{enumerate}
    \item In Section\;\ref{sec:profile_comparison} we compare the amplitude of the $\Delta\Sigma$ profile found from the DLIS high cluster sample to the expected sample in a suite of cosmological simulations. This constrains \oas.
    \item In Section\;\ref{sec:fits} we then fit a 3D density profile model to the observed $\Delta\Sigma$ profile for each cluster sample, deriving estimates of the splashback radii.
    \item In Section\;\ref{sec:radii_constraints} we explore how two features of the infall region, the splashback radius and the truncation radius in Eq.\;(\ref{eq:rho_model}), could be used to constrain cosmology.
\end{enumerate}

\subsection{Cosmological constraints from profile comparisons}
\label{sec:profile_comparison}

\begin{figure}
    \centering
    \includegraphics[width=\linewidth]{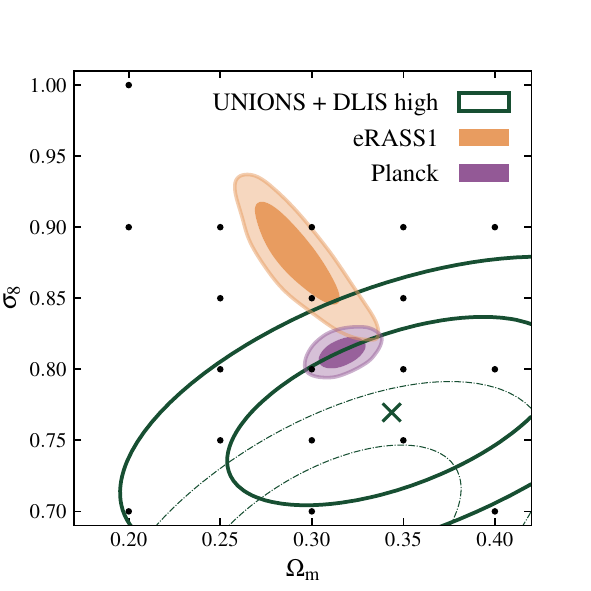}
    \caption{Cosmological constraints from comparing simulated dark matter halos, with UNIONS weak lensing profiles of DESI Legacy Imaging Survey (DLIS) galaxy clusters with $M_{500{\rm c}}>10^{14}M_{\odot}/h$. The observed profile in the infall region is compared to the profile in each of a suite of $19$ cosmological simulations varying $\Omega_{\rm m}$ and $\sigma_8$ (black points) and a goodness-of-fit $\chi^2$ value is calculated for each. A 3D paraboloid is fit to these $\chi^2$ values. The best fit cosmology is indicated with a green cross, $68.3\%$ and $95.5\%$ confidence regions at $\Delta\chi^2=2.3,6.81$ from the minimum are overlaid. The dash-dot contour is the case when no redshift re-weighting of the simulation profiles is included.}
    \label{fig:cosmo_from_profiles}
\end{figure}

Of our four cluster samples, DLIS high is the closest to being  complete, as argued by \cite{DESI_Legacy}. Thus, we will compare the mean projected mass profile for this sample to the simulated profiles in order to constrain cosmological parameters. We use the DLIS high sample with $M_{500{\rm c}}>10^{14} M_{\odot}/h$ as this produces a signal comparable to that expected in simulations, as seen in Fig.\;\ref{fig:obs_profiles}, whereas at lower masses the DLIS sample may be contaminated by projected systems with overestimated masses. Of course we can not be sure we have removed all interlopers from the sample, and there may still be a bias in the observed $\Delta\Sigma$ profile. In creating the matched simulation profiles, we convert the simulated halo masses from $M_{200{\rm c}}$ to $M_{500{\rm c}}$ by adopting the cosmological parameters of the particular simulation, and using the NFW concentration parameter determined by AHF. The converted masses have been mass-scattered, such that we are including the effect of low-mass systems entering the sample, and high mass systems leaving the sample. We then take the mean profile of all simulated halos with $M_{500{\rm c}}>10^{14}M_{\odot}/h$. For the fiducial cosmology result given by a solid green line in Fig\;\ref{fig:cosmo_from_profiles}, this mean includes redshift based weights to correct for the volume-limited nature of the observed catalogue and the large difference in the redshift distributions, seen in Fig\;\ref{fig:Mz}. The dash-dot green line shows the case where no such weight is included in the simulation profiles.

We use the results in Eq.\;(14) of \cite{Mpetha} to define the radial range of the infall region over which to compare the simulated and observed profiles. Adopting the constants associated with UNIONS gives an angular radius at which we expect the signal in the infall region to peak of $\theta_{\rm peak}=0.254\,$deg. Considering Fig.\;4 of \cite{Mpetha}, we use an angular range of log$(\theta_{\rm peak})\pm0.25\,$dex. To convert the angular range to a radial range, we use the mean redshift of the DLIS high sample and adopt a fiducial value of $\Omega_{\rm m}=0.3$. This translates into a radial range of $1.77-5.61\,$cMpc$/h$. This range is shown as the grey band in Fig.\;\ref{fig:obs_profiles}.

In each cosmological simulation with a given value of \oas, the amplitude of the simulated profile in the defined radial range is compared to the amplitude of the observed profile in the same range. We always compare observed and simulated profiles generated with the same value of $\Omega_{\rm m}$. Each simulation in Fig.\;\ref{fig:cosmo_from_profiles}, represented by a black dot, has an associated $\chi^2$ of the difference between the simulated and observed profiles in the infall region. We fit a 3D paraboloid to the $\chi^2$ values, and plot the minimum of the paraboloid with a green cross; this point has a reduced $\chi^2$ of $\chi_r^2=0.51$. Finding the contour with $\Delta\chi^2=1$ above the minimum gives marginalised $68.3\%$ constraints of $\Omega_{\rm m}=0.34\pm0.06$ and $\sigma_8=0.77\pm0.04$. To draw the 2D $68.3\%$ and $95.5\%$ confidence regions we plot contours of $\Delta\chi^2=2.3$ and $\Delta\chi^2=6.81$ from the minimum. Also plotted are \textsc{GetDist} \citep{getdist} chains from \textit{Planck} \citep{Planck} and the first eROSITA All-Sky Survey (eRASS1) \citep{eROSITA_abundance}. It should be noted that the paraboloid fit to the $\chi^2$ surface is itself uncertain. Drawing from the correlated best-fit parameters for the 3D paraboloid produces a set of possible paraboloids, leading to a variation in the reported constraints of $\pm0.02$.

\subsection{Fits to observed weak lensing profiles}
\label{sec:fits}

\begin{figure*}
    \centering
    \includegraphics[width=\linewidth]{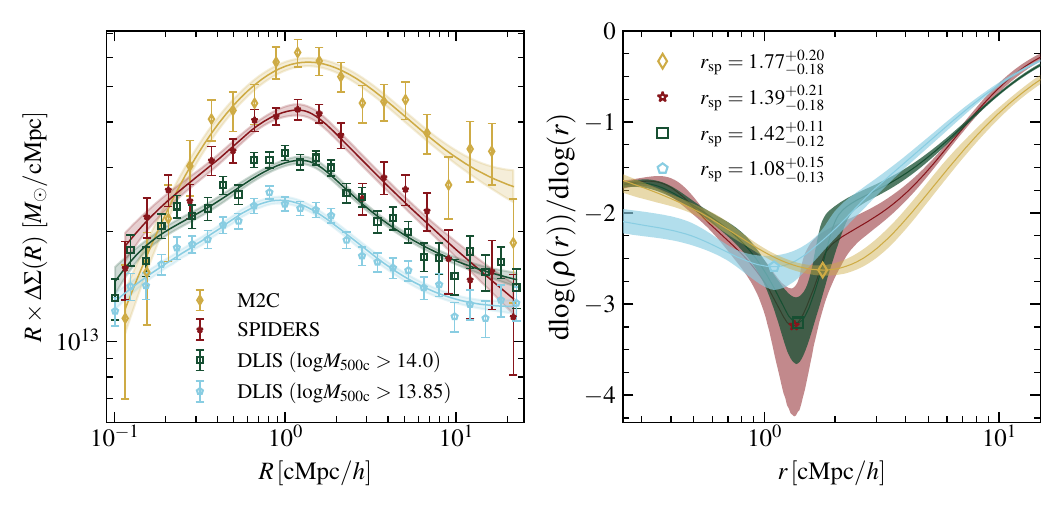}
    \caption{\textit{Left:} Observed excess surface mass densities for each of the cluster samples. For DLIS, we show the result from two different choices for the lower mass limit. These profiles are fitted with a 3D density profile model \citep{Diemer_fit}. \textit{Right:} The gradient of the 3D density profile fit is used to identify the splashback radius of each cluster sample.}
    \label{fig:fits}
\end{figure*}

\begin{figure*}
    \centering
    \includegraphics[width=\linewidth]{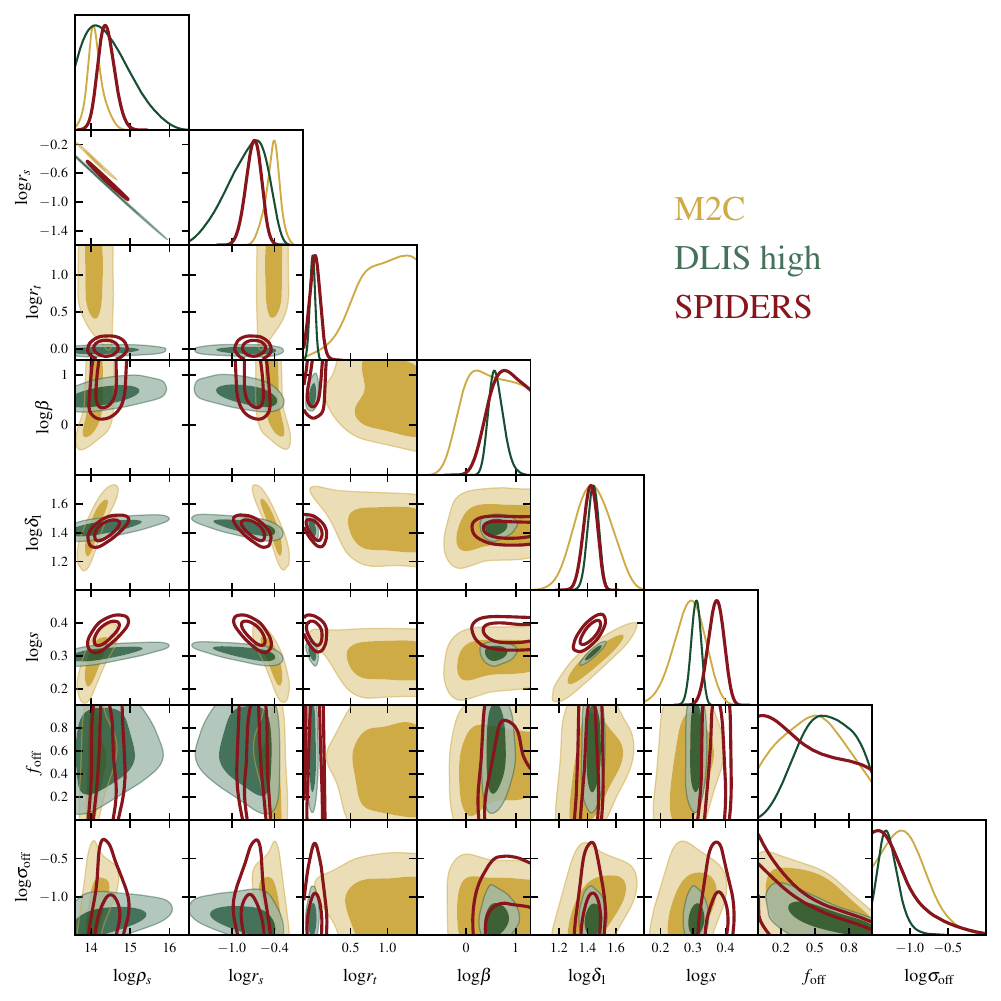}
    \caption{$68.3\%$ and $95.5\%$ credible regions for parameters of the 3D density profile model of \protect\cite{Diemer_fit}, fitted to the observed $\Delta\Sigma$ profile of three publicly available cluster catalogues. Chains are created using a Preconditioned Monte Carlo algorithm \citep{pocomc2,pocomc1}}
    \label{fig:corner}
\end{figure*}

The left panel of Fig.\;\ref{fig:fits} shows fits to mean mass profiles determined from weak lensing (in comoving units) in the four cluster samples. The fits differ from the simulation profiles shown in Fig.\;\ref{fig:obs_profiles} at large $R$ due to projection effects---the simulation profiles are found using the 3D density profile in spherical shells. The fits have a reduced $\chi^2$ of 1.09, 0.45, 0.71 and 0.5 for DLIS high, DLIS, SPIDERS and M2C respectively. The M2C sample has the largest amplitude, as it has the largest mean mass. It also has a slightly lower mean redshift than the SPIDERS sample. The DLIS sample has both the lowest mean mass and the highest mean redshift. Resulting fit parameters are shown in Table\;\ref{tab:fits}, and the $68.3\%$ and $95.5\%$ error contours derived from the corresponding PMC chain are shown in Fig\;\ref{fig:corner}. The right panel of Fig.\;\ref{fig:fits} shows the logarithmic gradient of the density profile from the fitting procedure with the splashback radius indicated. We make the simplifying assumption that the splashback radius, $r_{\rm sp},$ is the point where the gradient of the density profile is steepest, though it has been shown these two features are not always equivalent \citep{Diemer3_Halo}. 

From the best-fit density profile, we can derive $r_{200{\rm c}}$ and $M_{200{\rm c}}$ by locating the radius at which the average density within the halo, 
\begin{equation}
    \langle\rho(<r')\rangle = \frac{3\int_0^{r'} r^2 \rho(r) dr}{r'^3} \, ,
\end{equation}
reaches $200\rho_{\rm c}$. Because our cluster samples are mass-limited and each cluster mass is uncertain to $\sim\!0.2\,$dex, lower-mass systems are scattered into the sample, and clusters with masses in the mass-cut will be scattered out of the sample. This `Eddington bias' leads to a  sample with a true mean mass smaller than observed. `Observed mass' here refers to the masses estimated through some noisy observational cluster mass proxy, such as X-ray luminosity or stellar mass, with an intrinsic scatter. We include the Eddington bias in the simulation profiles in Fig.\;\ref{fig:obs_profiles}, finding it leads to an observed mean mass $\sim\!25\%$ larger than the true mean mass of the sample. When comparing our lensing-based masses in Table\;\ref{tab:fits} to the reported masses in Table\;\ref{tab:lenses}, we find the reported masses are a factor of $1.4\pm0.2$, $2.4\pm0.2$ and $2.1^{+0.3}_{-0.2}$ larger than the lensing masses for M2C, DLIS high and SPIDERS respectively. Therefore, the mass-scattering effect accounts for some, but likely not all, of the discrepancy in masses. Other selection effects could be biasing the reported masses in Table\;\ref{tab:lenses}. 

There are other reasons to be cautious about the masses reported from our fitting procedure in Table\;\ref{tab:fits}. There is a large degeneracy between parameters of the inner profile and mis-centering parameters. Under-estimating the degree of mis-centering can lead to an underestimation of the mass, which is likely to be occurring in the SPIDERS and DLIS high samples, indicated by the difference between the simulation and observed profiles in Fig\;\ref{fig:obs_profiles}. Another reason for the discrepancy in mass estimates could be the effect of baryonic processes on the density profile. In Fig.\;1 of \cite{Castro21}, for example, the authors find a $\sim\!10\%$ difference in $M_{200{\rm c}}$ when comparing hydrodynamical and dark-matter only simulations. For these reasons, we derive the fiducial cosmological constraints assuming the masses reported in Table\;\ref{tab:lenses} are correct, correcting only for Eddington bias (which is included in the simulations when creating matched simulation profiles and performing the profile comparison in Section\;\ref{sec:profile_comparison}). We could instead create matched simulation profiles assuming our lower lensing-derived masses are correct. Doing this shifts the contour in Fig.\;\ref{fig:cosmo_from_profiles}, and leads to constraints of $\Omega_{\rm m}=0.42\pm0.09$, $\sigma_8=0.82\pm0.05$. These results are consistent within $1\sigma$, but suggest uncertainties in cluster masses could increase our systematic errors by $\sim\!~50\%$, if not addressed in future follow-up.

Fig.\;\ref{fig:corner} shows how well the parameters of the infall region, notably the truncation radius $r_t$, can be determined for the SPIDERS and DLIS high samples. The observed M2C $\Delta\Sigma$ profile has the largest uncertainties due to the smallest number of clusters, and thus the fit parameters have the largest uncertainties in this case. All samples have a large fraction of sources with an offset, $f_{\rm off}$, but with a relatively small mis-centering amplitude of $\sigma_{\rm off} \simeq 40-65\,$kpc$/h$. These parameters are poorly constrained, however, having a large degeneracy with parameters of the inner profile. As M2C is constructed from purely X-ray and SZ selected clusters, which contain larger centering uncertainties, it is not surprising it has the largest reported degree of mis-centering.

A comparison between the case where log$\,\alpha$ and log$\,\delta_{\rm max}$ are fixed and left free in the fitting procedure is shown in Appendix\;\ref{app:fixed_params}.

\begin{table}
    \centering
    \begin{tabular}{c|lll}
 & M2C & DLIS high & SPIDERS \\
\hline
${\rm log}\,\rho_s$ & $14.11^{+0.17}_{-0.19}$ & $14.48^{+0.59}_{-0.6}$ & $14.40 \pm 0.21$ \\
${\rm log}\,r_s$ & $-0.43^{+0.1}_{-0.09}$ & $-0.81^{+0.3}_{-0.29}$ & $-0.69 \pm 0.11$ \\
${\rm log}\,\alpha$ (fixed) & $-0.49$ & $-1.36$ & $-0.80$ \\
${\rm log}\,r_t$ & $0.89^{+0.34}_{-0.36}$ & $-0.02^{+0.04}_{-0.03}$ & $0.02 \pm 0.07$ \\
${\rm log}\,\beta$ & $0.48^{+0.52}_{-0.54}$ & $0.61 \pm 0.15$ & $0.79^{+0.33}_{-0.34}$ \\
${\rm log}\,\delta_1$ & $1.44 \pm 0.13$ & $1.44 \pm 0.04$ & $1.42 \pm 0.05$ \\
${\rm log}\,\delta_{\rm max}$ (fixed) & $2.39$ & $2.78$ & $2.70$ \\
${\rm log}\,s$ & $0.29 \pm 0.05$ & $0.31 \pm 0.01$ & $0.37 \pm 0.02$ \\
$f_{\rm off}$ & $0.48 \pm 0.29$ & $0.59^{+0.24}_{-0.25}$ & $0.44^{+0.32}_{-0.36}$ \\
${\rm log}\,\sigma_{\rm off}$ & $-1.20^{+0.43}_{-0.37}$ & $-1.30^{+0.13}_{-0.15}$ & $-1.38^{+0.41}_{-0.37}$ \\
\hline
$r_{\rm sp}$ & $1.77^{+0.20}_{-0.18}$ & $1.42^{+0.11}_{-0.12}$ & $1.39^{+0.21}_{-0.18}$\\
$r_{200{\rm c}}$ & $0.99\pm0.05$ & $0.73^{+0.02}_{-0.01}$ & $0.84^{+0.03}_{-0.04}$ \\
log$(M_{200{\rm c}})$ & $14.37^{+0.07}_{-0.06}$ &  $13.95^{+0.04}_{-0.03}$ & $14.15^{+0.05}_{-0.06}$
\end{tabular}
    \caption{Parameter values from the profile fits for each of the three cluster samples. Profiles, in comoving units, were created using $\Omega_{\rm m}=0.3$ to convert angles and redshifts to distances. DLIS high uses a mass limit of $M_{500{\rm c}}>10^{14}M_{\odot}/h$. We also include derived results for the splashback radius $r_{\rm sp}$, found as the radius of the steepest slope of the logarithmic density profile, the mass $M_{200{\rm c}}$ and radius $r_{200{\rm c}}$. In all cases the $68.3\%$ credible region is quoted. $M_{200{\rm c}}$ has units of $M_{\odot}/h$, $\rho_s$ has units of $h^2 M_{\odot}/{\rm cMpc}^3$, while $r_s$, $r_t$, $\sigma_{\rm off}$, $r_{\rm sp}$ and $r_{200{\rm c}}$ have units of cMpc$/h$.}
    \label{tab:fits}
\end{table}

\subsection{Cosmological constraints using characteristic radii}
\label{sec:radii_constraints}

In Section\;\ref{sec:profile_comparison}, we derived the first cosmological constraints based on the amplitude of the shear signal in the infall region. As explored in \cite{Roan,Mpetha}, specific features of the infall region provide another possible test of cosmology. Extracting specific features, as opposed to comparing the profile over the entire infall region, may introduce model dependence and reduce the constraining power, but should produce results less dependent on vertical shifts of the shear profile. Here we explore two features, the splashback radius, and a specific feature of the density profile model in Eq.\;(\ref{eq:rho_model}), the truncation radius $r_t$. In both cases we find that the number of clusters in our three samples are not sufficient to obtain useful cosmological constraints.

\subsubsection{Splashback radius}

\begin{figure*}
    \centering
    \includegraphics[width=\linewidth]{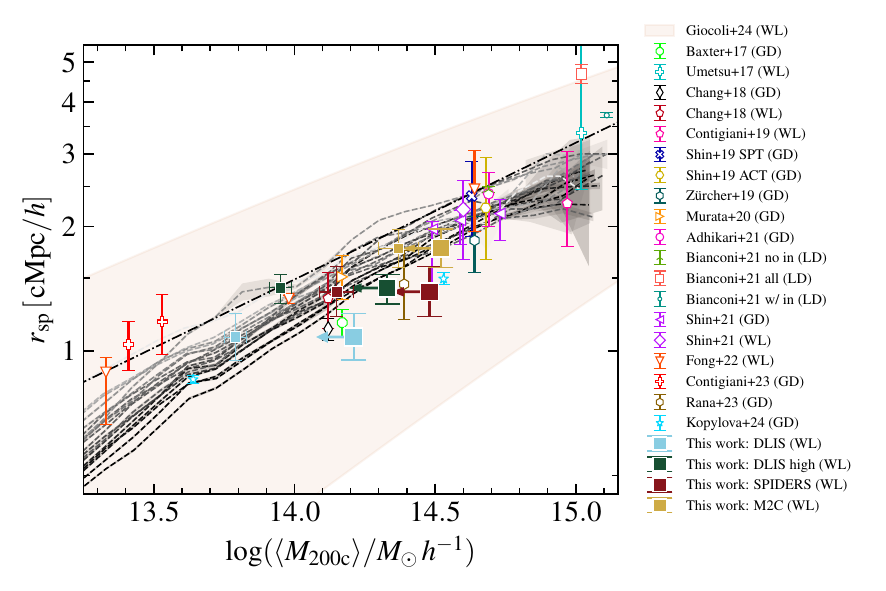}
    \caption{Dashed lines show the splashback radius $r_{\rm sp}$ as a function of mass in the $19$ cosmological simulations used in this work. Darker line colour corresponds to higher $S_8$ for that simulation, ranging from $0.57$ when $(\Omega_{\rm m},\sigma_8)=(0.2, 0.7)$, to $1.04$ with $(0.4,0.9)$. Uncertainties in the mean are given by shaded bands. The black dash-dot line is $2r_{200{\rm c}}$. Also overlaid are splashback radii from the literature, where mean masses of the cluster samples have been converted into $M_{200{\rm c}}$, and splashback radii into comoving units. These splashback radii have been measured from cluster profiles created by different methods: galaxy number density profiles (GD), weak lensing profiles (WL), and luminosity density profiles (LD). The four $r_{\rm sp}$ values from this work are included, either plotted with the reported catalogue masses (large filled squares), or our derived WL masses with uncertainties in Table\;\ref{tab:fits} (small filled squares). The arrows on the values from this work plotted using catalogue masses give an approximate indication of the masses with the Eddington bias removed. The orange shaded band is the fitting function, with uncertainty, from \citet{sp_obs_16}. In \protect\cite{sp_obs_12}, the authors use three versions of their catalogue to derive a splashback radius: including all clusters (`all'), only including clusters which contain infalling groups (`w/ in'), and only including clusters without any infalling groups (`no in'). }
    \label{fig:rsp}
\end{figure*}

\begin{figure}
    \centering
    \includegraphics[width=\linewidth]{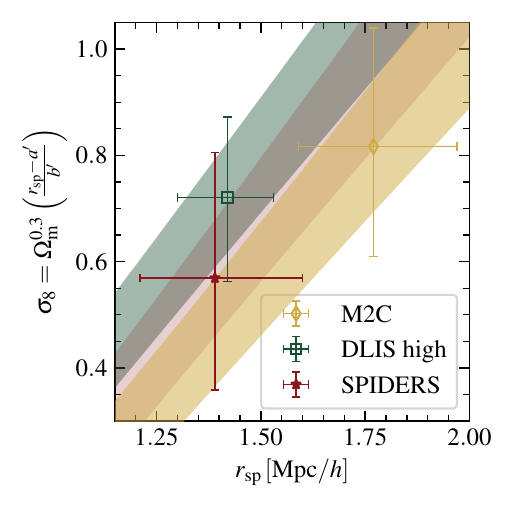}
    \caption{The shaded bands are the predicted scaling relation between the splashback radius $r_{\rm sp}$ and $\sigma_8$ in Eq.\;\eqref{eq:rspomsig8}, with a fixed $\Omega_{\rm m}=0.3$. The value for $a'$ and $b'$ in the scaling relations are derived from the simulation profiles created to match the three cluster catalogues used in this work. Also plotted is the observed $r_{\rm sp}$ from each catalogue, with the derived $\sigma_8$.}
    \label{fig:rsp_sigma8}
\end{figure}

Fig.\;\ref{fig:rsp} shows previously measured splashback radii from the literature, as a function of the mean mass $M_{200{\rm c}}$ of the cluster sample. We also include our own values from DLIS, SPIDERS, and M2C, measured from the UNIONS weak lensing profiles. We plot these measurements either using the reported catalogue masses (larger filled squares), or our weak lensing derived masses in Table\;\ref{tab:fits} (smaller filled squares). The grey lines are the simulation results, with darker colours indicating higher values of $S_8$. The uncertainty in the simulation results, indicated by a shaded band, is generally small except at large masses where sample variance is significant. It can be seen that several literature values, including our own, are low compared to expectations from simulations. However this is an unfair comparison, due to the mass-scatter effect described in Section\;\ref{sec:simulation_comparison}. We expect there to be more low-mass systems in the observed catalogue, meaning we expect $r_{\rm sp}$ to be smaller. The arrows on values from this work show the mean mass when a $25\%$ Eddington bias, found in our simulation profiles, is removed. A useful test is to compare the fitted values to the splashback radii in the matched simulation profiles in Fig.\;\ref{fig:obs_profiles} with $\Omega_{\rm m}=0.35$ and $\sigma_8=0.75$ (the closest cosmology to the result of our analysis in Section\;\ref{sec:profile_comparison}), as those profiles include the Eddington bias effect. When doing so, there is excellent agreement for DLIS high where the fitted value is larger than the simulation profile value by $0.07\sigma$. For SPIDERS, the fitted value is $0.64\sigma$ smaller than the simulation prediction. While not significant, this difference could be contributed to by the difference in the large-scale environment seen in Fig.\;\ref{fig:obs_profiles}. For M2C the fitted value is $0.70\sigma$ larger than the simulation prediction; as expected for a dynamically relaxed sample \citep{Diemer2_cosmo}, evidenced by the excess signal at large radii in Fig.\;\ref{fig:obs_profiles}.

The literature splashback radii in Fig.\;\ref{fig:rsp} have been calculated assuming different values of $\Omega_{\rm m}$, and the underlying cluster catalogues may also be subject to selection effects. Given these uncertainties, we do not attempt joint cosmological constraints on \oas\ using the splashback radius, but note that this will be possible with future data \citep{Mpetha}.

In Eq.\;(4) of \cite{Roan}, the authors present a relationship between the splashback radius, \oas, 
\begin{equation}
    r_{\rm sp} = a' + b' \left(\frac{\sigma_8}{\Omega_{\rm m}^{0.3}}\right) \, ,
    \label{eq:rspomsig8}
\end{equation}
calibrated using the same cosmological simulation suite as used in this work, but considering halos at $z=0$ only and with $M_{200{\rm c}}>10^{14}M_{\odot}/h$. We repeat their method using the matched simulation profiles (three of which are shown for each catalogue in Fig.\;\ref{fig:obs_profiles}), leading to values of $a' = 0.70 \pm 0.08$, $b' = 0.70 \pm 0.07$ for DLIS high; $a' =  0.82 \pm 0.09$, $b' = 0.70 \pm 0.08$ for SPIDERS; and $a' = 0.89 \pm 0.10$, $b' = 0.76 \pm 0.08$ for M2C. Using the value $\Omega_{\rm m}=0.3$ assumed in computing splashback radii, the associated $\sigma_8$ can be predicted. This gives $\sigma_8 = 0.72^{+0.16}_{-0.15}$, $\sigma_8 = 0.57^{+0.21}_{-0.24}$, and $\sigma_8 = 0.82^{+0.21}_{-0.22}$ for DLIS high, SPIDERS and M2C respectively. This information is summarised in Fig.\;\ref{fig:rsp_sigma8}. The shaded bands show the predicted scaling relations between $r_{\rm sp}$ and $\sigma_8$ based on the matched simulation profiles for each catalogue, with a fixed $\Omega_{\rm m}=0.3$. Assuming a larger $\Omega_{\rm m}$ would steepen this relation. The observed $r_{\rm sp}$ and derived $\sigma_8$, with uncertainties, are also plotted. While not yet competitive with other tests in the literature, the result from each catalogue is consistent with the results of Section\;\ref{sec:profile_comparison}.

\subsubsection{Truncation radius}

\begin{figure}
    \centering
    \includegraphics[width=\linewidth]{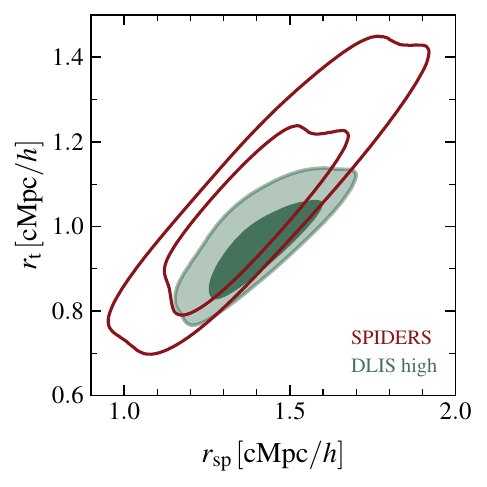}
    \caption{The best-fit splashback radius and truncation radius found from fits to UNIONS weak lensing profiles of two galaxy cluster catalogues.}
    \label{fig:rsp-rt}
\end{figure}

As pointed out in \cite{Diemer25}, the truncation parameter $r_{\rm t}$ of the density profile model in Eq.\;(\ref{eq:rho_model}) could provide a more robust tracer of the accretion history of clusters than the steepest slope of the density profile. Furthermore, the inferred value of the point of steepest gradient is very sensitive to the fitting function/algorithm used \citep{Diemer3_Halo}, whereas $r_t$ is shown to be a more stable indicator of accretion rate \citep{Diemer25}. Fig.\;\ref{fig:rsp-rt} shows the relationship between the inferred splashback radius and the truncation radius.

We check our consistency with the results of \cite{Diemer25} using their derived relation between $r_t/r_{200{\rm m}}$ and the peak height $\nu_{200{\rm m}}$,
\begin{equation}
    r_t/r_{200{\rm m}}=1.4-0.21\nu_{200{\rm m}} \, . \label{eq:rtr200m}
\end{equation}
We adopt the value of $r_{200{\rm m}}$ derived from our density profile fits, and convert the corresponding fitted $M_{200{\rm m}}$ to peak height utilising the \textsc{colossus} code-base \citep{colossus}. Using Eq.\;\eqref{eq:rtr200m}, we find predicted values of $r_t/r_{200{\rm m}}=0.88,0.90$ for SPIDERS and DLIS high respectively.  The corresponding fitted values for $r_t/r_{200{\rm m}}$ are $0.71^{+0.14}_{-0.12}$ and $0.75^{+0.11}_{-0.10}$. Considering uncertainty in the derived relation has not been included, which the spread in Fig.\;5 of \cite{Diemer25} indicates will not be negligible, this near agreement is encouraging and provides strong evidence for an empirical relation between the truncation radius and peak height.

The value of $r_{\rm t}$ in Table\;\ref{tab:fits} can be compared to the values from the matched simulation profiles in Fig.\;\ref{fig:obs_profiles}, over the whole grid of \oas. Doing so provides cosmological constraints consistent with those in Section\;\ref{sec:profile_comparison}, but with larger errors and greater dependence on systematics, particularly in the profile fitting, so we do not report these results. 

\section{Discussion}
\label{sec:discussion}

The mean formation time of dark matter halos depends on the values of cosmological parameters. Measuring the mass distribution in the infall region of galaxy clusters provides a novel method to estimate their average formation time. In cosmologies with low values of $\Omega_{\rm m}$ and high values of $\sigma_8$, clusters form earlier. This means they have more time to accrete material and become dynamically relaxed, leading to a more extended profile, and more mass in the infall region. In this work, we have used weak lensing profiles of cluster samples drawn from three publicly available catalogues to measure the mass distribution in the infall region, thereby constraining the background cosmology in which these clusters formed. The dependence of cluster formation time on cosmology is nearly orthogonal to the dependence of the halo mass function on cosmology. Thus, combining these two observables in a galaxy cluster catalogue can break the degeneracy in \oas\ without requiring external data sets.

We also measured the mean splashback radius for the same samples. In principle, the splashback radius can be a test of cosmology. The test is more complicated, however, since the value of $\Omega_{\rm m}$ used to convert angles and redshifts to distances when creating the $\Delta\Sigma(R)$ profile should match the value assumed in the simulation. Furthermore, \cite{Diemer25} has shown that the relationship between the 3D splashback radius $r_{\rm sp}$ and the steepest slope of the projected density profile, normally used to identify the splashback radius in observations, is in fact complex and model-dependent.

The infall region has the unique advantage that it is not sensitive to poorly modelled baryonic physics, meaning dark-matter-only simulations are sufficient to develop this novel cosmological probe. There are several major challenges when using the infall region to constrain cosmology (though these challenges are largely the same as those faced in cluster abundance studies). Incompleteness and selection effects in the cluster samples may bias the mean profiles, while projection effects may dilute the cluster sample with lower-mass systems. Despite the challenges, measuring the mass distribution of the infall region seems a promising approach. Given a reliable fitting function that maps from features of the infall region to cosmology, measurements of galaxy cluster profiles can be highly complementary to cluster abundance measurements. A reliable mapping from cosmology to features of the infall region will be the subject of a follow-up work.

\section{Conclusions}
 \label{sec:conclusions}
 
The publicly available cluster catalogues used in this work have some disadvantages for our method. For the SPIDERS and M2C catalogues, the small numbers and possible selection biases produce significant uncertainties in the mean profiles. For the DLIS catalogue, cluster redshifts and membership are estimated from photometric redshifts, and so projection effects may add many false detections, strongly diluting the signal. Applying a higher mass cut reduces the impact of spurious systems, but also reduces the number of clusters considerably, and with it the constraining power of the sample. Though the higher mass cut produces a mean projected mass profile more consistent with the simulations, it is possible false detections remain in the sample. The redshift distribution in the DLIS sample also differs from that of the simulations. While it was found in \cite{Mpetha} that the infall region varies weakly with redshift in comoving units, a more rigorous analysis would match the redshift distributions of simulated and observed clusters, instead of performing a simple re-weighting to estimate the effect of a matched redshift distribution. Despite these shortcomings, our method produces constraints of $\Omega_{\rm m}=0.34\pm0.06$ and $\sigma_8=0.77\pm0.04$. Allowing for unresolved systematic errors in cluster mass estimates increases these uncertainties by $\sim\!50\%$, so this remains an area for further work. The final UNIONS shape catalogue will cover a significantly larger area, and will also include photometric redshifts for each source, producing a more accurate boost correction. Thus, we expect our results to improve in the near future.

We also investigated using characteristic radii of the infall region as a cosmological test, finding that current measurements are not precise or accurate enough for meaningful constraints. We measured splashback radii of $1.39^{+0.21}_{-0.18}\,$cMpc$/h$, $1.77^{+0.2}_{-0.18}\,$cMpc$/h$ and $1.42^{+.11}_{-0.12}\,$cMpc$/h$ for SPIDERS, M2C and DLIS high respectively, noting that while these values are low compared to predictions from simulations, the differences are below $1\sigma$ when correcting for the Eddington bias effect. 

There is good potential for the infall region using forthcoming data. For it to reach its potential, the response of the infall region to cosmological parameters needs to be investigated in a grid of cosmologies that is both more finely resolved, and spans a greater number of parameters. Such simulation suites already exist: \textsc{AbacusSummit} \citep{abacus} and \textsc{DarkQuest} \citep{DarkQuest}. Their box size and resolution will provide two orders of magnitude more dark matter halos than used in this work, giving excellent prospects for a detailed theoretical investigation of the infall region, which we will perform in a follow-up work.

To forecast future potential for this test, we can assume all of the DLIS sources in the current UNIONS footprint with $M_{200{\rm c}} > 10^{14}M_{\odot}/h$ will have reliable mass and redshift measurements. This produces a sample of $\sim\!60,\!000$ clusters, a factor $\sim\!8$ increase in the number used in Fig.\;\ref{fig:cosmo_from_profiles}. Assuming the same UNIONS shape catalogue, the possible improvement is $\sqrt{8}$, leading to a $68.3\%$ contour area equal to that of eROSITA cluster abundances \citep{eROSITA_abundance}. The orthogonal degeneracy direction in the $\Omega_{\rm m}-\sigma_8$ plane means combining these results leads to a factor two reduction in constraints from cluster abundances alone.

\section*{Acknowledgements}

We thank members of the UNIONS collaboration, in particular Qinxun Li, Jack Elvin-Poole and Hunter Martin for useful discussions. We also thank Benedikt Diemer, Hironao Miyatake, Takahiro Nishimichi and Shogo Masaki for valuable input, Charles Kirkpatrick for providing the SPIDERS catalogue masses and uncertainties, and Vittori Ghirardini for providing the eROSITA \textsc{GetDist} chain. C. T. M. is funded by a Leverhulme Study Abroad Scholarship. J. E. T. and M. J. H. acknowledge support from the Natural Sciences and Engineering Research Council of Canada (NSERC), through Discovery Grants. H. Hildebrandt is supported by a DFG Heisenberg grant (Hi 1495/5-1), the DFG Collaborative Research Center SFB1491, an ERC Consolidator Grant (No. 770935), and the DLR project 50QE2305. This research was enabled in part by support provided by Compute Ontario (www.computeontario.ca) and the Digital Research Alliance of Canada (alliancecan.ca). The python packages \textsc{numpy}, \textsc{scipy}, \textsc{GetDist}, \textsc{matplotlib}, \textsc{lmfit}, \textsc{colossus}, \textsc{pocoMC}, \textsc{dsigma}, \textsc{camb}, \textsc{astropy} and \textsc{multiprocess} have been used in this work. This research has made use of the M2C Galaxy Cluster Database, constructed as part of the ERC project M2C (The Most Massive Clusters across cosmic time, ERC-Adv grant No. 340519). For the purpose of open access, the authors have applied a Creative Commons Attribution (CC BY) licence to any Author Accepted Manuscript version arising from this submission.

We are honored and grateful for the opportunity of observing the Universe from Maunakea and Haleakala, which both have cultural, historical and natural significance in Hawaii. This work is based on data obtained as part of the Canada-France Imaging Survey, a CFHT large program of the National Research Council of Canada and the French Centre National de la Recherche Scientifique. Based on observations obtained with MegaPrime/MegaCam, a joint project of CFHT and CEA Saclay, at the Canada-France-Hawaii Telescope (CFHT) which is operated by the National Research Council (NRC) of Canada, the Institut National des Science de l’Univers (INSU) of the Centre National de la Recherche Scientifique (CNRS) of France, and the University of Hawaii. This research used the facilities of the Canadian Astronomy Data Centre operated by the National Research Council of Canada with the support of the Canadian Space Agency. This research is based in part on data collected at Subaru Telescope, which is operated by the National Astronomical Observatory of Japan.
Pan-STARRS is a project of the Institute for Astronomy of the University of Hawaii, and is supported by the NASA SSO Near Earth Observation Program under grants 80NSSC18K0971, NNX14AM74G, NNX12AR65G, NNX13AQ47G, NNX08AR22G, 80NSSC21K1572 and by the State of Hawaii.

\section*{Data Availability}

The simulations used in this article will be shared on reasonable request to the corresponding author.

As for UNIONS, a subset of the raw data are publicly available via the Canadian Astronomical Data Center at http://www.cadc-ccda.hia-iha.nrc-cnrc.gc.ca/en/megapipe/. The remaining raw data and all processed data are available to members of the Canadian and French communities via reasonable requests to the principal investigators of the Canada-France Imaging Survey, Alan McConnachie and Jean-Charles Cuillandre. All data will be publicly available to the international community at the end of the proprietary period.


\bibliographystyle{mnras}
\bibliography{UNIONS_infall} 



\appendix

\section{Sky distribution}
\label{app:sky}

\begin{figure*}
    \centering
    \includegraphics[width=\linewidth]{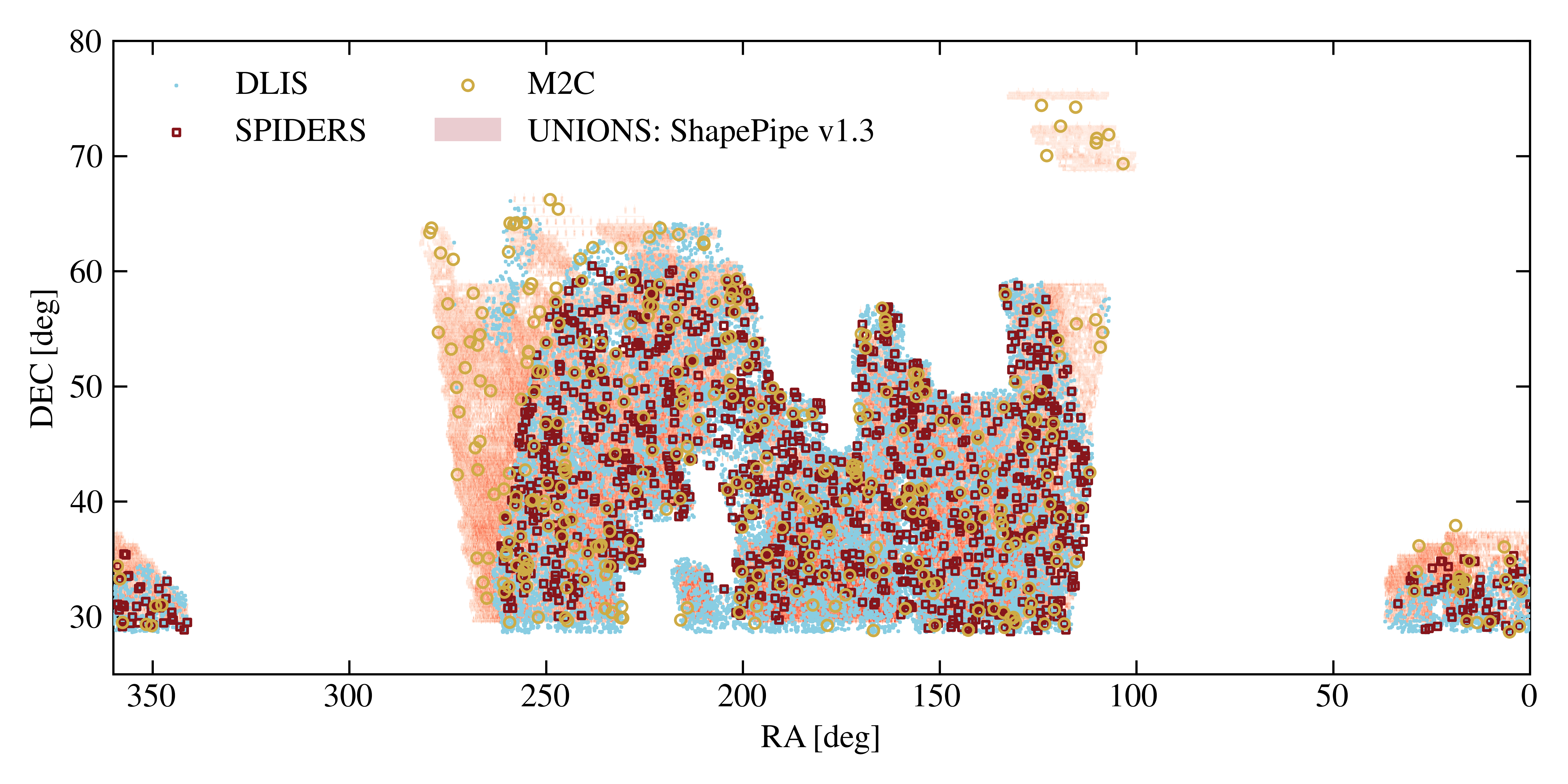}
    \caption{The footprint of the shape catalogue used in this work, UNIONS ShapePipe v1.3, overlaid with the positions of clusters lying in this footprint from the three publicly available catalogues used in this work.}
    \label{fig:sky}
\end{figure*}

Fig.\;\ref{fig:sky} shows the footprint of the UNIONS ShapePipe v1.3 catalogue, overlaid with the positions of all clusters from three publicly available catalogues used in this work.

\section{Boost factor correction}
\label{sec:boost}

The size of the boost factor correction in Eq.\;(\ref{eq:boost}) is shown for each cluster sample used in this work. It varies with radius, and impacts the amplitude of the profile in the infall region up to $5-10\%$.

\begin{figure}
\centering
\includegraphics[width=\linewidth]{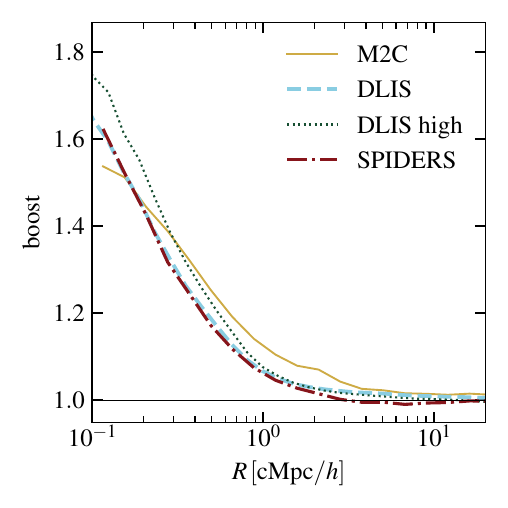}
    \caption{Boost factor calculated in \textsc{dsigma} for the observed $\Delta\Sigma$ profile (with $\Omega_{\rm m} =0.3$) of each publicly available cluster catalogue used in this work.}
    \label{fig:boost}
\end{figure}

\section{Reduced shear approximation}
\label{app:kappa}

\begin{figure}
    \centering
    \includegraphics[width=\linewidth]{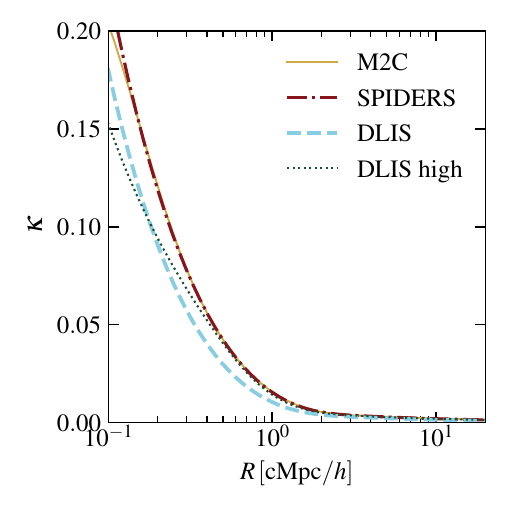}
    \caption{Convergence profiles for the three cluster catalogues used in this work.}
    \label{fig:kappa}
\end{figure}

In this work we assume the reduced shear is equal to the shear, ignoring a $1-\kappa$ correction. Here we justify that choice by estimating the size of $\kappa$ as a function of radius for the three cluster samples M2C, SPIDERS and DLIS high. The quantity to estimate is
\begin{equation}
    \kappa = \frac{\Sigma(R)}{\Sigma_{\rm crit}} \, .
\end{equation}
We can take $\Sigma_{\rm crit}$ from the lensing calculation, using the average value over all lenses in the sample. The $\Sigma(R)$ profile found when creating the matched simulation profiles in Fig.\;\ref{fig:obs_profiles} is used. The result is seen in Fig.\;\ref{fig:kappa}. The convergence correction drops below $20\%$ at $R=0.1\,$Mpc$/h$, where our fits to the profile begin, and is negligible in the infall region. We conclude that the reduced shear approximation is sufficient in our analysis.

\section{Fixing density profile model parameters}
\label{app:fixed_params}

\begin{figure}
    \centering
    \includegraphics[width=\linewidth]{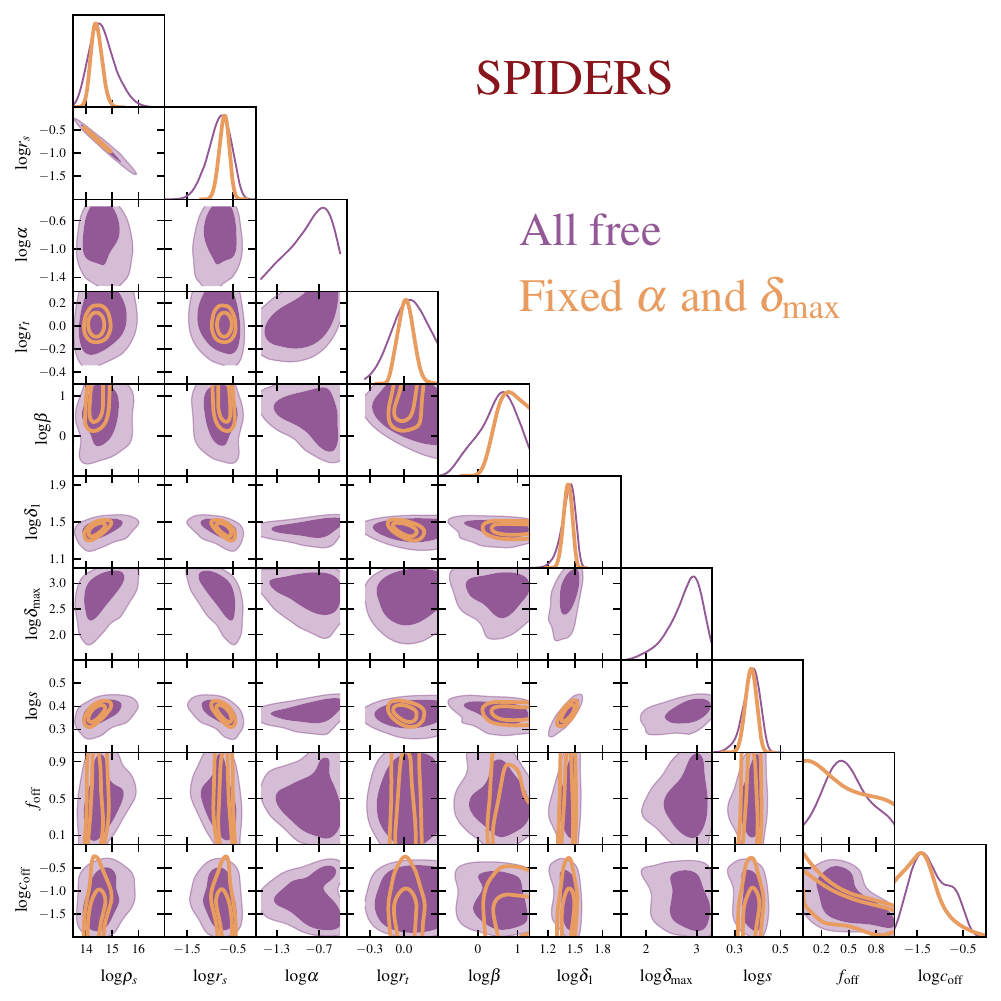}
    \caption{Comparison between best-fit parameters for the SPIDERS profile (only using clusters with $M_{500{\rm c}}>10^{14}M_{\odot}/h$), when all parameters are left free, or $\alpha$ and $\delta_{\rm max}$ are kept fixed to their best fit values from the all free case.}
    \label{fig:fix}
\end{figure}

Fig.\;\ref{fig:fix} demonstrates that the fit results for the density profile are largely the same when the parameters $\log\alpha$ and $\log\delta_{\rm max}$ are fixed or left free, but posteriors from infall region  parameters are reduced.

\bsp	
\label{lastpage}
\end{document}